\documentstyle[agupp,psfig]{article} 
\lefthead{HELMSTETTER et al.}
\righthead{Diffusion of Aftershocks}	

\newcommand{\be}{\begin{equation}}
\newcommand{\ee}{\end{equation}}
\newcommand{\ba}{\begin{eqnarray}}
\newcommand{\ea}{\end{eqnarray}}

\paperid{2003JB*****}
\authoraddr{Agn\`es Helmstetter, Institute of
Geophysics and Planetary Physics, University of California, 
Los Angeles, California. (e-mail: helmstet@moho.ess.ucla.edu)}
\authoraddr{Guy Ouillon, Laboratoire de Physique de la
Mati\`{e}re Condens\'{e}e,  CNRS UMR 6622 Universit\'{e} de
Nice-Sophia Antipolis, Parc Valrose, 06108 Nice, France
(e-mail: Ouillon@aol.com).}
\authoraddr{Didier Sornette, Laboratoire de Physique de la
Mati\`{e}re Condens\'{e}e,  CNRS UMR 6622 Universit\'{e} de
Nice-Sophia Antipolis, Parc Valrose,06108 Nice, France,
and Department of Earth and Space Sciences and Institute of
Geophysics and Planetary Physics, University of California, Los Angeles,
California. (e-mail: sornette@naxos.unice.fr)}
\input{update.tex}
\begin{document}
 
\title{Are Aftershocks of Large Californian Earthquakes Diffusing?}

\author{Agn\`es  Helmstetter$^1$, Guy Ouillon$^2$ and Didier Sornette$^{1-3}$}
\affil{$^1$ Institute of Geophysics and Planetary Physics,
University of California, Los Angeles, California 90095}
\affil{$^2$ Laboratoire de Physique de la Mati\`ere Condens\'ee, CNRS UMR6622
and Universit\'e des Sciences, Parc Valrose, 06108 Nice Cedex 2, France.}
\affil{$^3$ Department of Earth and Space Sciences,
University of California, Los Angeles, California 90095}

\begin{abstract}
We analyze 21 aftershock sequences of California to test for
evidence of space-time diffusion. 
Aftershock diffusion may result from stress 
diffusion and is also predicted by any mechanism of
stress weakening. Here, we test an alternative
mechanism to explain aftershock diffusion, based on multiple
cascades of triggering.
In order to characterize aftershock diffusion, we develop two methods,
one based on a suitable time and space windowing, the other using a wavelet
transform adapted to the removal of background seismicity. Both
methods confirm that diffusion of seismic activity is very weak,
much weaker than reported in previous studies.
A possible mechanism explaining the weakness of observed diffusion
is the effect of geometry, including the
localization of aftershocks on a fractal fault network
and the impact of extended rupture lengths which control
the typical distances of interaction between earthquakes.

\end{abstract}

\begin{article}

\section{Introduction}

Aftershocks are defined by their clustering properties both in time
and space. The temporal clustering obeys the
well established (modified) Omori law [{\it Utsu et al.}, 1995]
which states that the rate of aftershocks decays as
\begin{equation}
n(t) = \frac{A}{(t+c)^p}~,
\label{jslls}
\end{equation}
where the Omori exponent $p$ is generally found between $0.5$ and $2$.
In expression (\ref{jslls}),
the origin of time is the occurrence of the main event. The
offset time $c$ is often not well-constrained and may partly
account for the incompleteness of catalogs close to mainshocks.
It provides a regularization at short times preventing the event rate
to reach infinity at the time of the mainshock. This law with 
the cut-off $c$ may be derived from a variety of physical mechanism
(e.g. [{\it Dieterich}, 1994]).

The spatial organization of aftershocks is more complex and less understood.
On the one-hand, it is recognized that many aftershocks occur
right on the fault plane or in its immediate
vicinity (at the scale of the mainshock rupture
length) and this is used for 3D visualization of the rupture
surface (see for instance {\it Fukuyama} [1991]).
The observation that the spatial distribution of earthquakes in
California is characterized by a fractal dimension close to $2.2$
[{\it Kagan and Knopoff}, 1980] and the fact that this measure is
dominated by aftershock
clustering suggests a rather complex network of active faults
[{\it Sornette}, 1991].
The clustering of aftershocks on or close to the mainshock rupture
probably reflects local stress concentration at asperities that
served to stop the rupture and lock the fault. On the other hand,
aftershocks also occur away from fault ruptures,
due to various triggering mechanism, the simplest one
being stress transfer and increasing
Coulomb stress [{\it King et al.}, 1994; {\it Stein et al.}, 1994;
{\it Stein}, 1999; {\it Toda et al.}, 2002].

Both temporal and spatial properties argue for the presence of
triggering processes, which could also be expected to lead to
combined space and time dependence of the organization of aftershocks.
Indeed, several studies have reported
so-called ``aftershock diffusion,'' the phenomenon of
expansion or migration of aftershock zone with time
[{\it Mogi}, 1968; {\it Imoto}, 1981; {\it Chatelain et al.}, 1983;
{\it Tajima and Kanamori}, 1985a,b; {\it Ouchi and  Uekawa}, 1986;
{\it Wesson}, 1987; {\it Rydelek and Sacks}, 1990; {\it Noir et al.},
1997; {\it Jacques et al.}, 1999].
Immediately after the mainshock occurrence, most aftershocks are
located close to the rupture plane of the mainshock, then aftershocks
may in some cases migrate away from the mainshock,
at velocity ranging from $1$ km/h to $1$ km/year 
[{\it Jacques et al.}, 1999; {\it Rydelek and Sacks}, 2001].
This expansion is not universally observed, but is more important in 
some areas than in others [{\it Tajima and Kanamori}, 1985a].
Most of these studies are qualitative and are based on a few sequences
at most. {\it Marsan et al.} [1999, 2000] and {\it Marsan and Bean} [2003]
have developed what is to our knowledge the first
systematic method to analyze space-time interactions between pairs
of earthquake and report evidence for strong diffusion (see our discussion
below on these results). A similar method has also been used by
{\it Huc and Main} [2003].
The present state of knowledge on aftershock diffusion
is confusing because contradictory results have
been obtained, some showing almost systematic diffusion whatever the
tectonic setting and in many areas in the world,
while others do not find evidences for aftershock diffusion 
[{\it Shaw}, 1993].
This may reflect the intrinsic variability in space and time of
genuine diffusion properties and also possible
biases of the analyses, for example, due to background seismicity.

 From a physical viewpoint, the diffusion of aftershocks is usually
interpreted as a diffusion of the stress induced by the mainshock,
either by a viscous relaxation process [{\it Rydelek and Sacks},  2001],
 or due to fluid transfer in the crust  [{\it Nur and Booker}, 1972; 
{\it Hudnut et al.}, 1989;  {\it Noir et al.}, 1997].
However, such a stress diffusion process is not necessary to
explain aftershock diffusion. Recent studies have indeed suggested
that aftershock diffusion may result from either a rate and state
friction model  [{\it Dieterich}, 1994; {\it Marsan et al.}, 2000]
or a sub-critical growth mechanism [{\it Huc and Main}, 2003], 
without invoking any process of stress
diffusion. Actually, aftershock diffusion is predicted by any model
that assumes that  (i) the time to failure increases if the applied
stress decreases and (ii) the stress change induced by the mainshock
decreases with the distance from the mainshock. Therefore, aftershocks
further away from the mainshock will occur later than those closer to the
mainshock, because the stress change is smaller and thus the failure
time is larger. An increase of time to failure with the applied stress
is predicted by many models of aftershocks, including rate and state
friction  [{\it Dieterich}, 1994], sub-critical crack growth 
[{\it Das and Scholz}, 1981; {\it Yamashita and  Knopoff}, 1987;
{\it Shaw}, 1993], damage or static fatigue laws [{\it Scholz}, 1968;
{\it Lee and Sornette}, 2000].
For instance, the sub-critical growth model of [{\it Das and Scholz}, 1981]
predicts that the time to failure $t_c$ of an aftershock decreases with the
stress change $d \sigma$ according to 
\be
t_c \sim d\sigma^{-n}~,
\label{tc}
\ee
where $n$ is the stress corrosion index.
If the stress change decreases with the distance $r$ from the
mainshock according to
\be
d\sigma(r)  \sim r^{-1/2}
\label{sigma}
\ee
in the near field (at distances  from the fault smaller than the
mainshock rupture length), then the average time to
failure increases with $r$ as
\be
t_c (r) \sim r^{n/2}
\label{tcr}
\ee
By inverting $r$ as a function of $t$,
the typical distance $R(t)$ of aftershocks occurring at time $t$
after the mainshock is thus expected to increase according to 
[{\it Huc and Main}, 2003]
\be
R(t) \sim t^{H}~,
\label{Rt}
\ee
with $H=2/n$.
Expression (\ref{Rt}) thus predicts a very slow sub-diffusion, with a
diffusion exponent $H=2/n \approx 0.06 $ for a corrosion index
close to 30 as generally observed [{\it Huc and Main}, 2003].
The diffusion exponent is even smaller at large distances from
the mainshock because the stress decreases as $ \sim 1/r^3$
and thus $H=1/(3n)$.
If the time to failure decreases exponentially with the applied
stress, as expected for example for static fatigue laws 
[{\it Scholz}, 1968],
the typical distance $R$ increases logarithmically with time.
Such a slow diffusion is difficult to observe in real data due to
the limited number of events and the presence of background seismicity.
Take $H=0.2$ for instance: one needs five decades in time range to observe one
decades in space range. The smallness of the diffusion exponent $H$
predicted by this and other models (see Appendix A) explains why it
may be difficult to observe it.

Another alternative explanation of aftershock diffusion, that does not
rely on any stress diffusion process, is multiple triggering.
{\it Ouchi and Uekawa} [1986] first reported
that the diffusion of aftershocks is often due to the occurrence
of large aftershocks and the ensuing localization of secondary
aftershocks close to them.
The apparent diffusion of the seismicity may thus result from a
cascade process; the mainshock triggers aftershocks which in turn
trigger their own aftershocks, and so on, thus leading to an
expansion of the aftershock zone.
A recent series of papers have investigated quantitatively this concept of
triggered seismicity [{\it Helmstetter and Sornette}, 2002a,b,c; 2003a;
{\it Helmstetter et al.}, 2003] 
in the context of the so-called epidemic-type aftershock (ETAS) model
of earthquakes seen as point-wise events, which was
introduced by {\it Ogata} [1988] and in a slightly different form 
by {\it Kagan and Knopoff} [1981; 1987].
The ETAS model assumes that each earthquake triggers aftershocks,
with a rate that decays in time according to Omori's law, and
which increases with the mainshock magnitude. The distribution of 
distances between triggered and triggering
earthquake is assumed to be independent of the time.
Using this model, {\it Helmstetter and Sornette} [2002b] showed that,
under the right conditions (see Appendix A),
the characteristic size $R$ of the aftershock cloud may increase
as a function of time $t$ since the mainshock according to
expression (\ref{Rt}), where $H$ is a function of both the exponent of
Omori's law and of the exponent describing the spatial
interactions between events (see Appendix A). 
Figure \ref{map} presents
results from numerical simulations of the ETAS model to show
how cascades of multiple triggering can produce aftershock diffusion.
The analysis of the ETAS model in  [{\it Helmstetter and Sornette}, 2002b]
offers some predictions that can be tested in real 
aftershock sequences. Diffusion should be observed 
only if the Omori exponent $p$ is smaller than 1, and
the diffusion exponent $H$ should decrease with $p$ if $p<1$.
In addition to the diffusion law  (\ref{Rt}), the ETAS 
model also predicts a decrease of the Omori exponent $p$
with the distance from the mainshock.

The fundamental difference between the diffusion predicted by the 
ETAS model and by models based on stress weakening mechanisms is that, in the
later, diffusion derives from the direct effect of the mainshock while, 
in the former, there is no diffusion of direct (first generation)
aftershocks and diffusion results from the cascade of secondary aftershocks.

Motivated by these different empirical and theoretical works,
we revisit the issue of the existence of aftershock diffusion
and of its characterization. For this,
we develop two different and complementary methods
to test these predictions. We analyze aftershock sequences in California
and interpret the results in the lights of the predictions obtained
from  the ETAS model [{\it Helmstetter and Sornette}, 2002b] 
(see Appendix A for a brief description of the relevant predictions).

The organization of the article is as follows.
We first discuss the problems encountered in previously
published analyses of diffusion in real data.
We then explain and use a direct method of analysis of
several aftershock sequences in California, followed
by a second method based on wavelet analysis. This second method aims
at optimizing the removal of the perturbing influence of the 
seismicity background.
We then discuss the limits of the theoretical and numerical analysis,
and interpret our empirical results in the light of various
available models.

Our analyses are carried out on
two different catalogs, (i) the catalog of Southern California seismicity
provided by the Southern California Seismic Network
for the period 1932-2000, and (ii) the catalog of  Northern 
California seismicity
provided by the Northern California Seismic Network since 1968.
The minimum magnitude for completeness  ranges from $M=3.5$ in 1932
to $M<2$ for the two catalogs since 1980. The average uncertainty on 
earthquake location
is about $1$ km for epicenters, but is larger for hypocenters.
Therefore we consider only the spatial distribution of epicenters.

\section{Methodological Issues  \label{secmetobsdif}}

A priori, the observed seismicity results from a complex interplay
between tectonic driving, fault interactions, different spatio-temporal
field organizations (stress, fluid, plastic deformation, phase
transformations, etc.) and physical processes of damage and rupture.
Notwithstanding this complexity,
from the viewpoint of empirical observations, one may
distinguish two classes of seismicity: 
the seismicity resulting from the tectonic loading which is often
taken as a constant source (uncorrelated seismicity) and the triggered 
seismicity resulting from earthquake interactions (correlated seismicity).
An alternative classification of seismicity defines earthquakes
as either foreshock, mainshock, aftershock or background seismicity.
The definition of aftershocks and foreshocks requires the specification
of a space-time window used to select foreshocks (resp. aftershocks)
before (resp. after) a larger event defined as the mainshock.
The background is the average level of seismicity prior to the mainshock or
at large times after the mainshock.
Of course, such classification is open to criticism in view of the
inescapable residual arbitrariness of the selection
of foreshocks, mainshocks and aftershocks
[{\it Helmstetter et al.}, 2003; {\it Helmstetter and Sornette}, 2003].
Nevertheless, it offers a useful reference
for testing what is measured in earthquake diffusion processes.

Following previous studies, our analyses presented below assume
that earthquake occurrence follows a point process,
because we deal with the space and time organization of aftershock
epicenters. Physically, this would be justified when all
spatial and temporal scales are larger than source rupture dimension
and duration. In reality, this is not the case and the point process
representation may lead to incomplete or misleading conclusions,
as we point out in section \ref{gemoakma}.

\subsection{Effect of uncorrelated seismicity}

The major problem when analyzing real seismicity data in search
of some evidence for diffusion comes from the mixture of 
correlated seismicity which can display diffusion
and of uncorrelated seismicity (noise).
The latter can significantly alter
the evaluation of the characteristic distance of the aftershock zone.
To illustrate this problem, we analyze in Figure
\ref{synthaft} a synthetic catalog generated by superposing a large number
of independent aftershock sequences, without adding a constant
seismicity background. 
This synthetic catalog has been generated by superposing 
1000 mainshocks with a Poissonian distribution
in time and space and with a Gutenberg-Richter distribution of magnitudes,
and by adding aftershocks sequences of each mainshock, with a number
of aftershock increasing as $10^{0.8 M}$  with the mainshock magnitude $M$.
If we study, as in previous studies,
the average distance between all pairs of events in the catalog as
a function of their average time difference, the superposition of uncorrelated
aftershock sequences induces an apparent and spurious diffusion, over almost
four orders of magnitude in time.
This apparent diffusion comes from the superposition of a large number
of aftershock sequences, with a power-law distribution of duration and
spatial extension, resulting from the increase of the number of
aftershocks and of the length of the aftershock zone with the
mainshock magnitude, together with a power-law distribution of
earthquakes sizes (Gutenberg-Richter law).

In order to study the diffusion of aftershocks, triggered directly
or indirectly by the same source, one strategy is
to study individual aftershock sequences of large earthquakes
selected by some clustering algorithm, in order
to remove the influence of uncorrelated seismicity. This is the
approach followed in the present paper.

\subsection{Discussion of the method of Marsan et al.}

\begin{figure}[t]
\begin{center}
\psfig{file=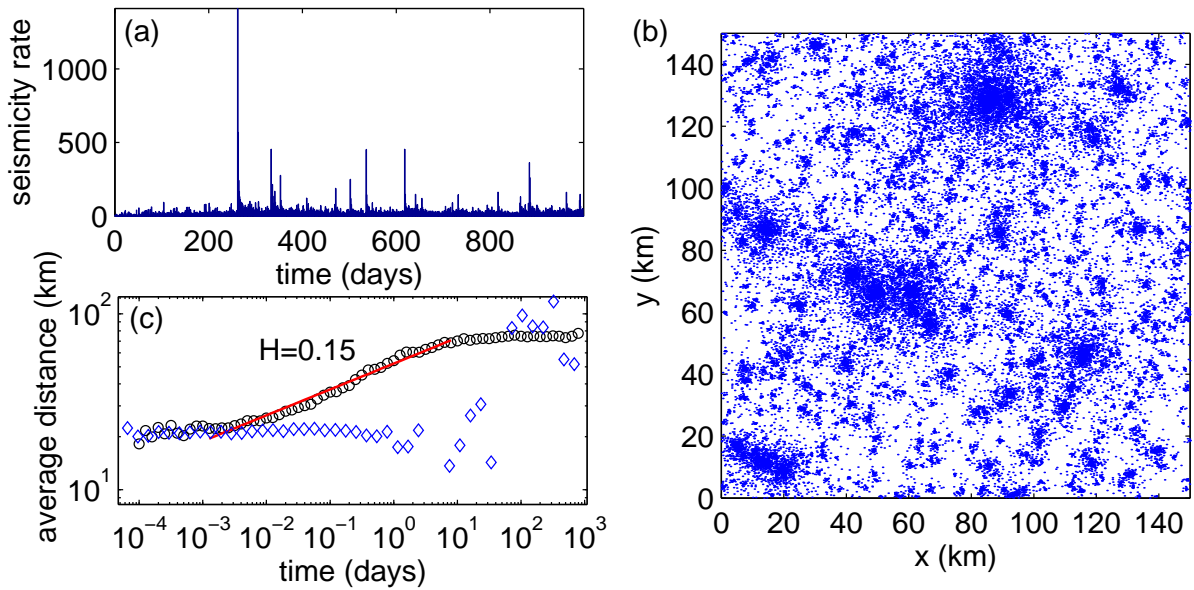,width=16cm}
\caption{\label{synthaft}
Analysis of a synthetic aftershock sequence.
We have built a synthetic catalog by superposing
a large number of independent aftershock sequences.
We have used 1000 mainshocks, with a Poissonian distribution
in time and space. Each mainshock generates only direct aftershocks
with a rate given by (\ref{nmgjedl}), with $K=10$, $\alpha=0.8$,
$\theta=-0.1$, $c=0.001$ day, $\mu=2$ and $d$ equal to the rupture
length of the mainshock. The distribution of distances between
aftershocks and mainshocks is thus independent of the time between
mainshock and aftershocks. The global number of events in the
catalog is 40000 including the 1000 mainshocks.
The seismicity rate (a) displays several peaks
corresponding to the occurrence of large mainshocks, as observed
for California seismicity. The map (b) shows large clusters
corresponding to the aftershock sequences of the largest mainshocks.
The average distance between all pairs of events (c) shown as circles
increases with the average time between events as $R \sim t^H$ with
$H=0.15$ (solid line), for a large interval of the time between
events $[0.001-10]$ days. For larger times, the average distance
saturates to $R \approx 80$ km, half the size of the catalog.
The diamonds show the results obtained with the method of {\it Marsan et
al.} [2000]. At early times ($t<1$ day), the average
distance is constant and the method is effective to remove the
influence of uncorrelated seismicity. But at large times when the
aftershock activity is small, there are
large fluctuations of the average distance, because the method is
very sensitive to the noise.}
\end{center}
\end{figure}

In contrast, Marsan et al. 
[{\it Marsan et al.}, 1999; 2000; {\it Marsan and Bean}, 2003]
have proposed a different
method, which uses all available seismicity and
considers all pairs of events independently of their magnitude,
with the advantage that (1) the
corresponding data set is much larger than for individual aftershock sequences
and (2) their method does not require the sometimes
arbitrary definition of mainshocks and aftershocks. In principle,
their approach provides a systematic search of possible space-time
correlations between earthquakes.
However, there is a potential problem with their
method stemming from the effect of
the background seismicity and of uncorrelated
seismicity, as illustrated in figure \ref{synthaft}.
Indeed, they study the average distance between all points as a function
of the time between them. Therefore, their count is performed over
a mixture of very different earthquake populations,
a significant fraction of events being probably unrelated causally.

In order to remove the influence of this uncorrelated seismicity, they
use the global catalog to estimate the average distance between two points,
and they remove the contribution of the average seismicity to estimate
the spatio-temporal distribution of the correlated seismicity.
Specifically, Marsan et al.
calculate a pair-wise time-dependent space-space
correlation function {\it corrected} for the background by subtracting
the long-time average value of this space-space correlation function,
from which they estimate an average distance between pairs as a 
function of time.
The growth of this average distance with time may then qualify a causal
dependence between earthquakes, through a diffusive process.
Marsan and others [{\it Marsan et al.}, 1999; 2000; 
{\it Marsan and Bean}, 2003] studied in this way several catalogs 
at different spatial scales, from the scale of a
mine to world-wide seismicity, and observed that the average
distance between two earthquakes increases as a power-law of the time
between them, with an exponent often close to $H=0.2$, indicative of a
sub-diffusion process. They interpreted their results as a mechanism
of stress diffusion, that may be due to fluid transfer with
heterogeneous permeability leading to sub-diffusion.

In this sense, Marsan et al. do not try
to establish a causal dependence between earthquakes but rather
to detect a correlation in their spatio-temporal
organization. Recall that correlation shows that two events
are related, but it does not determine their cause and effect relationship.
This is because there are basically three possible explanations
for the observation of correlation:
(i) The correlation is a coincidence; (ii) One event causes the other;
(iii) The two events are both caused by some third event.
In their detection of spatio-temporal correlations in seismicity, 
Marsan et al. do not distinguish between explanation (ii) nor (iii) while
in contrast the present paper emphasizes explanation (ii).

We have performed many tests of Marsan et al.'s method
on synthetic catalogs. A typical
test is shown as the diamond symbols in figure \ref{synthaft}.
At early times ($t<1$ day), the average
distance between earthquake pairs is constant, as it should be,
and the method removes adequately the
influence of uncorrelated seismicity. But at large times when
aftershock activity is small,
the average distance exhibits large fluctuations as a function of time.
This is due to the corrective term which becomes ill-conditioned at large
times and leads to a large sensitivity to noise and finite sample sizes.
We have found that, in most of our synthetic catalogs constructed without
genuine diffusion, the average distance obtained by Marsan et al.'s method
is approximately constant at early times as it should, but then 
crosses over to another noisy
plateau at long times, as a result of the ill-defined correction term.
This noisy crossover from a constant to another higher plateau
obviously gives an apparent growth of the average distance between pairs of
earthquakes as a function of time, which may compete with or hide a 
genuine signal.
It seems to us that several of the figures on the time dependence of the
average distance, which are presented in 
[{\it Marsan et al.}, 1999; 2000; {\it Marsan and Bean}, 2003],
exhibit this feature of a more or less
constant or very weak growth followed by a more abrupt jump to another
noisy plateau. If one interprets Marsan et al.'s results in terms
a causal diffusion processes,
the evidence on diffusion provided in
[{\it Marsan et al.}, 1999; 2000; {\it Marsan and Bean}, 2003]
is only suggestive and additional
tests of the methods on synthetic and real data should be performed
in order to understand its effects on the
analysis of real data and to improve the correction term.
Marsan et al.'s analysis is perhaps better
interpreted as evidence of a spatio-temporal correlation of seismicity
resulting from viscous relaxation, stress pulses and other processes
operating at large scales. But then, their analyses should not be
interpreted using models of causal diffusion based on rate- and 
state-dependent friction law [{\it Marsan et al.}, 2000] 
or on sub-critical crack  growth [{\it Huc and Main}, 2003], 
that model only mainshock-aftershock pairs and 
not arbitrary pairs of earthquakes.

\section{Windowing method}

\subsection{Method}

In this study, we analyze individual aftershock sequences and consider
the diffusion of the seismicity triggered directly or indirectly by 
the mainshock.
We adjust the values of the time window $T$ and the space window $D$ 
used to select aftershocks so that the rate of background activity is 
negligible in comparison with the aftershock rate. The background
seismicity rate is estimated by the average seismicity before the mainshock.
We also adjust the minimum magnitude $m_0$ and the minimum time $t_{min}$
after the mainshock in order to obtain a catalog that is complete for
$t_{min}<t<T$ and $m>m_0$.

In order to estimate the average size $R$ of the
aftershock area as a function of the time
from the mainshock, we define the barycenter of the
aftershock sequence as the reference point
because the mainshock epicenter has no reason a priori to
be the source of the aftershock sequence.
The average size of the aftershock area $R$ is obtained
as the average distance between the aftershocks and the barycenter.
We find that this procedure yields a diffusion exponent $H_r$ that
is always a little larger than the diffusion exponent estimated
from the average distance between the aftershocks and the mainshock
epicenter, as done in {\it Marsan et al.} [1999; 2000], 
{\it Marsan and Bean} [2003] and {\it Huc and Main} [2003].
We also introduce a refinement to take into account the anisotropy
of the aftershock zone and the spatial extension of the mainshock
rupture. For this, we compute the axes $a$ and $b$ of inertia of the
whole aftershock sequence as a function of the time after the mainshock,
in the spirit of {\it Kagan} [2002]. Geometrically, this corresponds
to approximating the map of aftershocks as filling an ellipse
with small axis $b$ and large axis $a$.
This provides two additional diffusion exponents $H_a$ and $H_b$.
For strike-slip mainshocks, $b(t)$ measures the average
distance between aftershocks and the fault plane,
while the large axis gives the average distance between
aftershocks and the barycenter on the fault plane.
 
We also measure the Omori exponent by plotting the rate of
aftershock activity as a function
of time in a log-log plot, and by measuring the slope $p$ by a linear 
regression
for $t_{min}<t<T$. We have also used a maximum likelihood method to estimate
both the $p$ and $c$ values of the modified Omori law. In most cases,
the two methods provide similar values of $p$.
We also estimate the variation of $p$ with the distance $r$ between
the mainshock and its aftershocks by selecting aftershocks at different
distances between the mainshock. As described in Appendix A, a
prediction of the ETAS model concerns the modification of the distribution
of distances $r$ between aftershocks and the mainshock epicenter
with time. We plot the distribution of distances $r$ between
the mainshock and its aftershocks for several time windows to test
if there is an expansion of the aftershock area with time.

We have tested this method using synthetic catalogs generated
with the ETAS model, including a constant seismicity background.
We have checked that our method provides a reliable estimate of the diffusion
exponent and is almost insensitive to the background activity as long as
the duration of the time window is sufficiently short so that the seismicity
is dominated by aftershocks.

\subsection{Results}

We have analyzed 21 aftershock sequences of major earthquakes in
California with number of aftershocks larger than 500, in addition
to two aftershock sequences of large earthquakes (Kern-County,
M=7.5, 1952, and San-Fernando, M=6.6, 1971) which have less than 500
events. The results for all these 21 sequences are listed in Table 
\ref{tabdifobs}.
The different values of $H$, measured either by the average distance
from the barycenter ($H_r$) or from the inertial axes ($H_a$ and $H_b$),
generally give similar results.
Exceptions are the Morgan Hill, Loma-Prieta, and Joshua-Tree events
which have $H_b > H_a$, and the Imperial Valley event which gives $H_b<H_a$.
The diffusion exponent $H$ is always positive, and generally smaller
than $0.1$. The estimated standard deviation on $H$, measured for
synthetic catalogs with the same number of events and similar
$p$-values, is about 0.05.
Thus, most aftershock sequences do not exhibit significant diffusion.

\begin{figure}
\begin{center}
\psfig{file=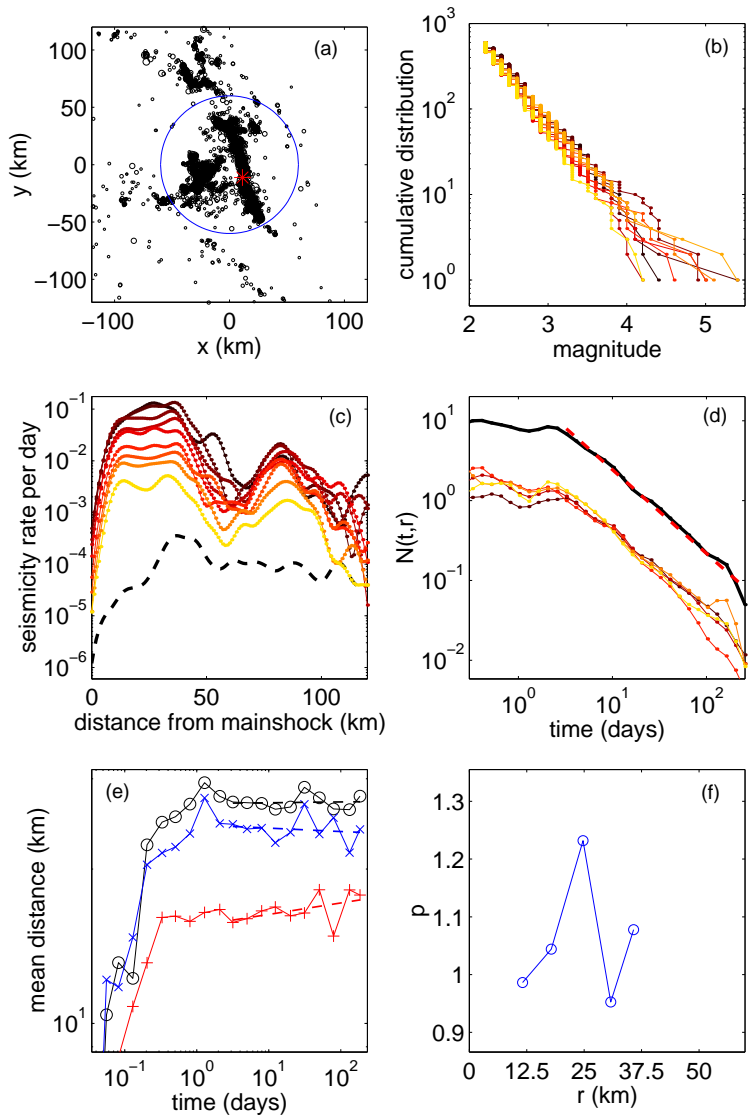,width=10cm}
\caption{ \label{landersdif} Analysis of the June, 28, 1992,
$M=7.3$ Landers aftershock sequence.
(a) Map of aftershocks, the mainshock epicenter is shown by a star.
(b) Magnitude distribution for different time windows (time increasing
from black to gray)  showing that the magnitude distribution
is stable over time, and that the catalog is complete above $m=2.2$
after 3 days after the mainshock.
(c) Rate of seismic activity as a function of the distance from the mainshock
for different times after the mainshock (increasing time from top to bottom
(black to gray)). The background activity preceding the mainshock is shown as a
dashed line.
(d) Rate of aftershocks for the whole sequence (solid black line at
the top) and fit by an Omori law (dashed gray line),
and  rate of aftershocks for different distances from the mainshock
(increasing distance from gray to black).
(e) Characteristic size of the aftershock cluster as a function of the
time from the mainshock, measured by the average distance from the
barycenter (circles), or from the small ('+') and large ('x') inertial
axes.
(f) Variation of the Omori exponent with the distance from the mainshock.}
\end{center}
\end{figure}

\begin{figure}
\begin{center}
\psfig{file=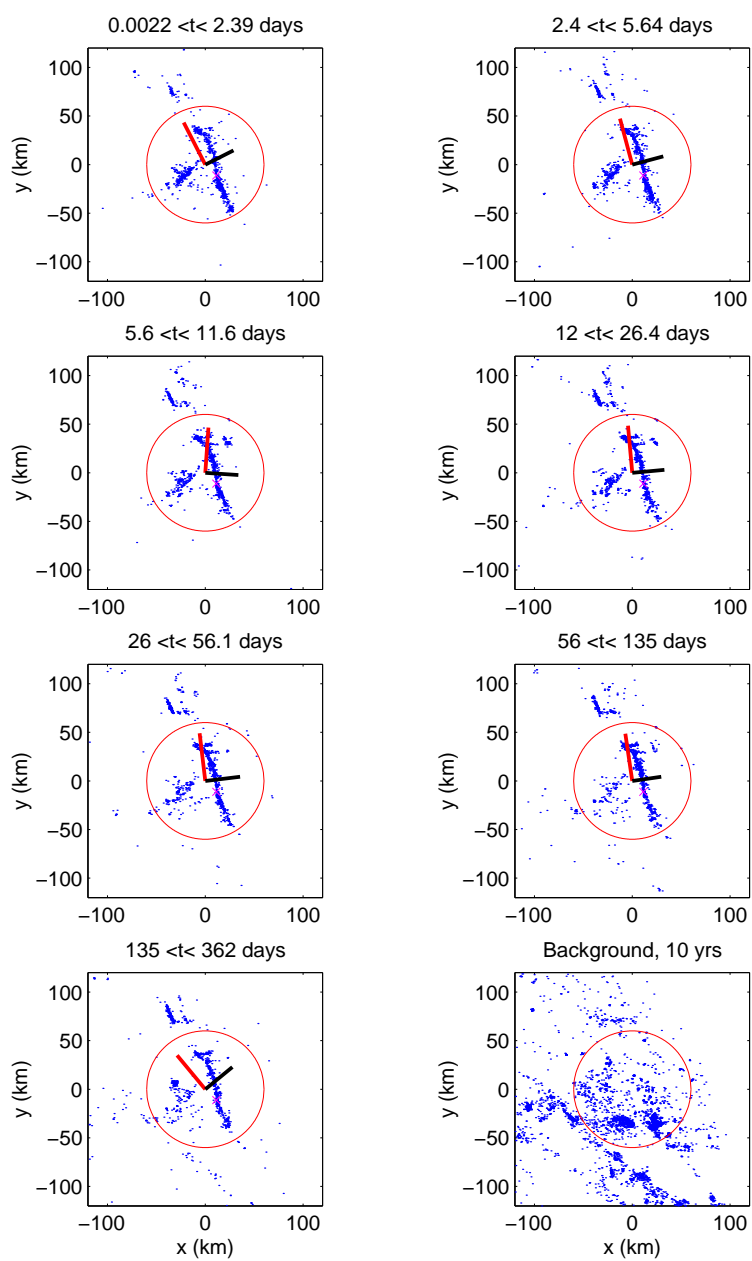,width=10cm}
\caption{ \label{landersmap}
Map of the aftershocks of the Landers earthquake, for different time windows
with 1000 events in each plot, showing the stationarity of the
spatial distribution of aftershocks. The epicenter is shown by a
cross. The gray and black lines show the large and small elliptical
axes respectively (multiplied by a factor 2 for clarity).}
\end{center}
\end{figure}

Details of the analysis for the aftershock sequence following the 1992
$M=7.3$ Landers event are shown in Figures \ref{landersdif} and
\ref{landersmap}. For this sequence, we measure an Omori exponent
$p=1.1$, which is stable when looking at different distances $r$.
The characteristic cluster size is also stable over more than two
orders of magnitude in time leading to $H \approx 0$. Similar results are
obtained for the elliptical axes $a$ and $b$.
The analysis of the distance distribution at different times also
confirms that there is no discernible diffusion of seismic activity.

\begin{figure}[t]
\begin{center}
\psfig{file=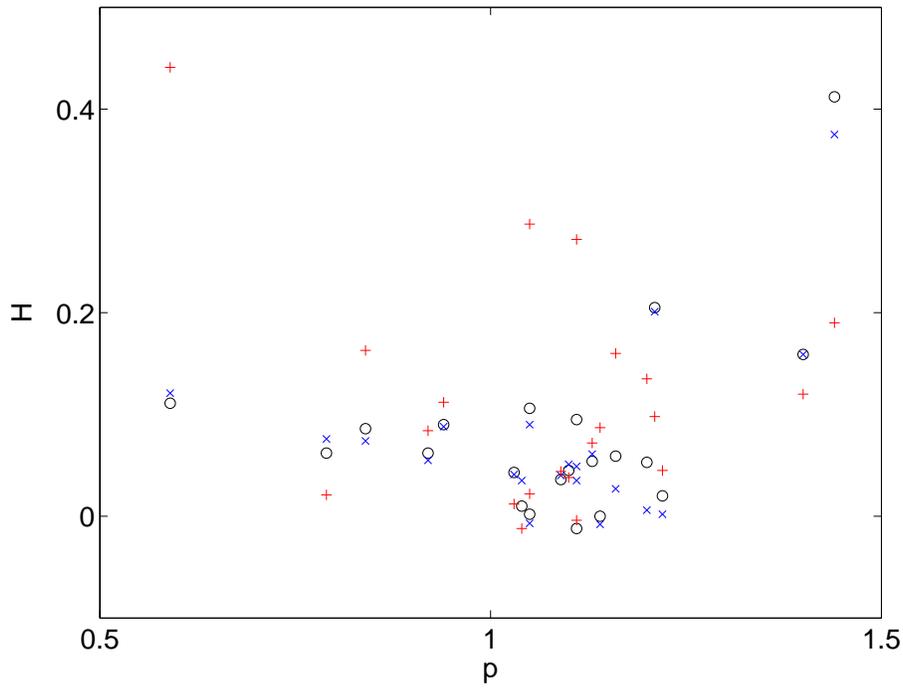,width=12cm}
\caption{ \label{pH} Diffusion exponents $H$ as a function of the Omori
exponent $p$ for the aftershock sequences described in Table
\ref{tabdifobs}. The circles give the diffusion exponent $H_r$
evaluated with the mean distance from the barycenter (radius of gyration), the
crosses correspond to the diffusion exponent $H_a$
and the '+' correspond to $H_b$.}
\end{center}
\end{figure}

Figure \ref{pH} summarizes the results for $p$ and $H$ listed in
Table \ref{tabdifobs}. The first result we should stress is that
all our values of the diffusion exponent $H$ are quite small when compared
with previous studies [{\it Marsan et al.}, 1999; 2000; 
{\it Marsan and Bean}, 2003]. For the reasons
explained in section \ref{secmetobsdif}, we believe
that Marsan et al.'s results may have been affected
by the background seismic activity,
and are quantifying different processes.
One can observe a weak negative correlation between $p$ and $H$ for $p<1$
but this negative correlation disappears when including data with $p>1$
which gives a very strong scatter with two among the largest $H$ obtained
for the largest $p>1$.

\section{Wavelet method}

The detection of aftershock diffusion may be biased
(i) by other (possibly) independent aftershock sequences and (ii)
by background seismicity.
In the windowing method, we have addressed these problems by
analyzing (a) individual aftershock sequences of large earthquakes
over a space-time window (b) with seismicity rate much larger than the
background estimated over a ten-year average.

We now present a second method based on wavelet
analysis, which introduces two innovations. First, it uses a smoothing kernel
(or mother wavelet) constructed in order to remove Poissonian background
seismicity. Second, it uses the expected scaling relationship relating space
and time that characterizes
diffusion to derive scaling laws obeyed by the wavelet coefficients. The
exponents $p$ of the observed Omori law and $H$ of aftershock diffusion
are then obtained by optimizing the compliance of
the wavelet coefficients to these scaling laws.
The wavelet approach displays the data all at once in the
space and time-scale dimensions in order to determine $p$ and $H$
simultaneously. In contrast, the windowing method displays the data
twice, once in the time dimension to obtain the exponent $p$ and a second
time to obtain $R(t)$ and the exponent $H$. The
wavelet method can thus be seen as a simultaneous two-dimensional
time scale-space determination of $p$ and $H$, in contrast with the 
previous windowing
method consisting in two independent one-dimensional
time series analyses, one for $p$ and the other for $H$.

\subsection{General diffusion scaling relation}
 
In the following, we will assume that a main event occurs
at time $t_0=0$, and is followed by a cascade of aftershocks,
superimposed on the long term background seismicity rate,
which is for now considered as a set of Poissonian,
independent events. We thus neglect the possible interactions between
the background noise and the cascade events, and that independent events
trigger their own cascades. We will come back to this point at the end of the
derivation and show that it is a reasonable assumption from a geophysical
point of view for short catalogs such as those we consider.
At any time $t$ following the mainshock,
the spatial and temporal evolution of the
seismicity rate $n(r,t)$ can be approximated by
\begin{equation}
n(r,t) ~dr ~dt = A(r,t)~ dr~ dt +  B(r,t) ~2\pi r~ dr ~dt~,
\label{ngjdl}
\end{equation}
where $r$ is the spatial distance from the
mainshock epicenter. $A(r,t)$ is the aftershock rate per unit time
and per unit distance from the mainshock:  $A(r,t) dr$ is thus
the number of events per unit time at time $t$
within an annulus of radii $r$ and $r+dr$.
In the following, a Poissonian distribution will
be assumed for the temporal structure of the background 
seismicity rate $B(r,t)$ per unit time and
surface and no restrictions will be imposed on its
spatial structure. This is an interesting aspect of the present
wavelet method.

The term $A(r,t)$ reflects the spatio-temporal structure of the
Omori law, describing the relaxation of the stress and of other 
physical fields,
which occurs in the vicinity of the main source and beyond. The
qualifying signature of diffusion is expressed by the following
general scaling relationship coupling time and space:
\begin{equation}
A(r,t) dr dt = Q {1 \over t^{p+H}} f\left(\frac{r}{Dt^H}\right) dr dt~,
\label{kgjle}
\end{equation}
where $H$ is the diffusion exponent, $p$ is the Omori law exponent
and the diffusion constant
$D$ is such that $Dt^H$ has the dimension of a length. $Q$ is a constant and
$f$ is an integrable function which depends on the physics of the diffusion
processes. The $1/t^{p+H}$ prefactor stems from the requirement that
the integral of $A(r,t) $ over all space should retrieve the
Omori law $\sim 1/t^p$.
For $H=0$, no diffusion occurs, while the value $H=1/2$
gives standard diffusion.
$H>1/2$ characterizes a superdiffusive regime and
$H<1/2$ corresponds to a subdiffusive regime which is the regime relevant
to aftershocks.

The problem is that the background term $ B(r,t) $ in (\ref{ngjdl}) is
not of the form (\ref{kgjle}) and therefore
may spoil the detection of diffusion. In other words,
the spatial and temporal structure of $B(r,t)$ scrambles the signal $A(r,t)$.
We now describe the wavelet approach that addresses this problem
by minimizing the impact of $B(r,t)$.

\subsection{Kernel smoothing in time}

Introducing the temporal kernel or mother wavelet $W(t)$, we consider
the wavelet transform of the signal $N(R,t)$ (where $N(R,t)$ is the spatial
integration of $n(r,t)$ for $r$ between $0$ and $R$) computed at time $t=0$
and time scale $a$ using the wavelet $W$
\begin{equation}
C_{a}(R) = \frac{1}{a} \int_{0}^{\infty} N(R,t) 
W\left(\frac{t}{a}\right) ~ dt~,
\label{mngmwsl}
\end{equation}
As each event can be considered
as a Dirac function in time, the integral in (\ref{mngmwsl}) reduces
to a discrete sum.
The scale factor $a$ allows us to dilate or contract the kernel $W$ in order
to get insight into the temporal structure
of $N(R,t)$ at various time scales $a$.

We use a kernel with zero average so that any
stationary process $S(R,t)$ uncorrelated in time does not contribute 
to $C_{a}(R)$:
\begin{equation}
\int_{0}^{\infty} S(R,t) W\left(\frac{t}{a}\right) \, dt = 0~.
\end{equation}
Here, $S(R,t)$ is the rate of background events occurring within a circle
of radius $R$ at time $t$.
This property allows us to get rid of the background seismicity without
presuming anything about its spatial structure, nor about the specific
time occurrence of such events. In this way,
background events are erased on average without needing to identify
them in the seismic catalog. This is an important quality of the
present wavelet analysis compared with previous studies of aftershocks.
Strictly speaking, for the background seismicity to disappear by this
procedure, we need (i) to consider large mainshocks, (ii) to have a small
probability that a large event due to the background is generated during
any aftershock sequence and (iii) to assume that the number of events
in the triggering cascade generated by any background event is low and
that all these events have small magnitudes.
These conditions ensure that the triggering cascades resulting from the
background $S(R,t)$ do not interact with the cascade of aftershocks
induced by the mainshock and that the duration of $S(R,t)$-induced
cascades is short compared with any time scale $a$ used in the wavelet
analysis. These conditions would be too restrictive if applied to the
whole span of a seismic catalog, but we expect them to be approximately
realized over the short time span of each aftershock sequence
whose duration is generally observed to be no more than
the order of weeks to years.

Now, taking the wavelet transform of $A(r,t)$ given by (\ref{kgjle}) gives
\begin{equation}
C_{a}(R) = Q D a^{-p} \int_{0}^{\infty}
\tau^{-p}~G\left(\frac{R/a^H}{D\tau^H}\right) ~W(\tau) ~ d\tau~,
\label{mgmjwlw}
\end{equation}
where $G$ is the integral of $f$ and $\tau=t/a$. Expression (\ref{mgmjwlw})
implies the scaling law
\begin{equation}
C_{a}(R) = a^{-p}~ C_{1}(R/a^H)~,  \label{mgmnsas}
\end{equation}
which relates the wavelet coefficient at time scale $a$ of the
time series of earthquake rate
within a circle of radius $R$ centered on the mainshock to the
wavelet coefficient for time scale $1$ and radius $R/a^H$, by a simple
normalization by $a^{-p}$. We will
refer hereafter to this scaling law as the $H$-scaling law.
This law (\ref{mgmnsas}) implies that, for any possible different values of
$a$ and $R$, plotting
$a^{p}C_{a}(R)$ as a function of $R/a^H$ leads to a collapse
of all points onto a single ``master'' curve.

Expression (\ref{mgmnsas}) can be transformed into
\begin{equation}
R^{\frac{p}{H}}C_{a}(R) = C_{aR^{-1/H}}(1)~, \label{gmjmwsw}
\end{equation}
which now provides a relationship between the wavelet
coefficient computed at time scale $a$ and radius $R$ and the
wavelet coefficient computed at time scale $aR^{-1/H}$ for a unit
radius. This scaling law (\ref{gmjmwsw}) will thereafter be referred to
as the $1/H$-scaling law. Expression (\ref{gmjmwsw}) shows that,
for any possible different values of $a$ and $R$, plotting
$R^{\frac{p}{H}}C_{a}(R)$ as a function of $aR^{-1/H}$ leads to a
collapse of all points onto another single ``master'' curve.

Both scaling laws (\ref{mgmnsas}) and (\ref{gmjmwsw}) capture
mathematically in a universal way the possible diffusion of aftershocks
around their mainshock. These scaling laws give access to the diffusion
exponent $H$ but do not say anything on the shape of the ``master'' curves,
which should derive from the specific properties
of $f(r/Dt^H)$ and of $W(t)$. The interest in using the two
scaling laws (\ref{mgmnsas}) and (\ref{gmjmwsw}) is that they magnify
and thus stress differently the small and long time and well as the
short and large distance part of the data.

The $1/H$-scaling law (\ref{gmjmwsw}) must have a
master curve with two asymptotes. If $R \gg Da^{H}$ (i.e., for small
normalized time scales compared to the radius),
it can easily be shown that the master curve is
a power-law of exponent $-p$. This result is
compatible with the exponent expected
for the whole sequence, that is, for the Omori law describing
the decay of the seismic rate of all events within a circle
of infinite radius $R$. If $R \ll D a^{H}$, then the
master curve is again a power law, but with
exponent $-(p+H)$. The computation of $C_{a}(R)$ for
small and large $R$'s
should thus provide $p$ and $H$. We will use below
a more sophisticated method that uses any range of
$R$ values.

\subsection{Choice of the smoothing kernel}

\begin{figure}[t]
\begin{center}
\psfig{file=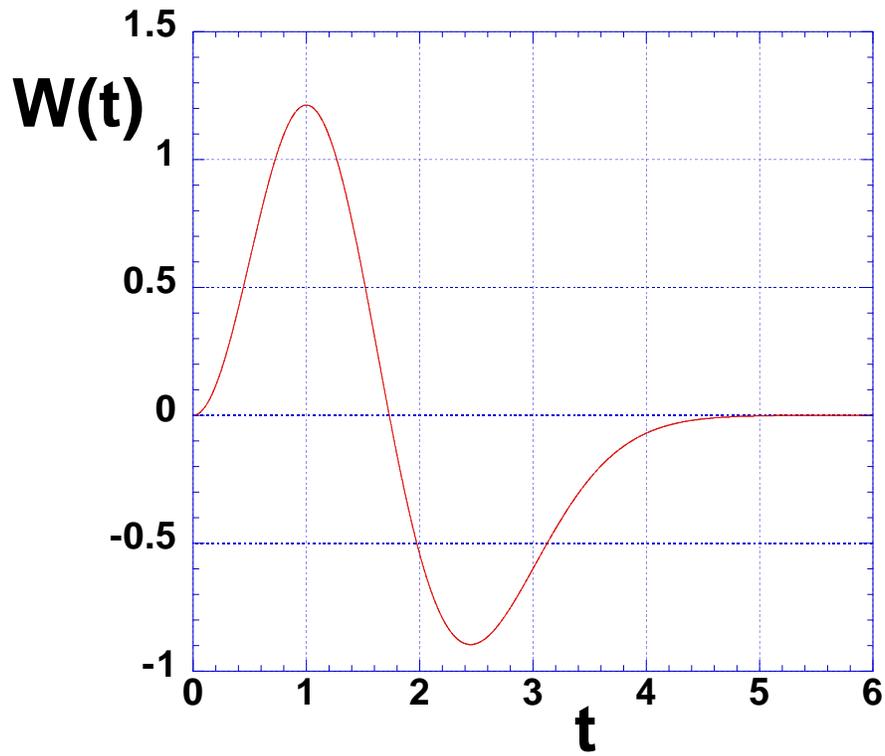,width=12cm}
\caption{ \label{kerwav} Wavelet kernel used for the wavelet
analysis of aftershock diffusion, defined by (\ref{mngnmlq}).
This wavelet kernel and its derivative both vanish at time $t=0$
and has a zero mean over the interval $[0, +\infty[$.}
\end{center}
\end{figure}

Our choice of the kernel or mother wavelet $W$ has been dictated
by the following considerations.
First, the modified Omori law (\ref{jslls}) as well as the ETAS model
(\ref{nmgjedl}) introduce a short-time cut-off $c$ that accounts for
seismicity just after the mainshock. Since this cut-off breaks down
the exact self-similarity of the Omori kernel, it leads to corrections
to the expected diffusion resulting from event triggering cascades
at short times [{\it Helmstetter and Sornette}, 2002b]. Since
$c$ is in general found small, the
seismicity rate is very large just after the mainshock which may lead
to finite-size and instrumental saturation effects just after the
mainshock. In addition, even if these limitations are removed
there may be so many aftershocks near $t=0$ that many of them are
generally interwoven in seismic
recordings, and most of them can not be interpreted and located
properly.
We thus construct a wavelet kernel $W$ which minimizes
the weight given to the seismicity rate at short times after mainshocks.
Specifically, we choose the wavelet kernel shown
in Figure \ref{kerwav}, which vanishes at $t=0$,
has also zero time derivative at the origin and has zero average:
\begin{equation}
W(t) = (3t^2-t^4)\exp(-\frac{t^2}{2})~.
\label{mngnmlq}
\end{equation}
This kernel looks like an aliased sine function. This wavelet
kernel is reasonably well localized both
in time and scale and its simple expression allows for fast computations
on large data sets. The time unit shown in this figure \ref{kerwav}
has no special meaning since our wavelet analysis is using
scale ratios rather than on absolute scales.
Note that the chosen wavelet kernel has properties
quite different from those, for instance,
of the classical Mexican hat wavelet, widely
used in the analysis of
singularities, which has a maximum amplitude at $t=0$, and would thus be
inappropriate for the reasons listed above.

This choice of the wavelet kernel leads to the decay of
the wavelet coefficients $C_{a}(R) \sim 1/a^{3}$ at large times after the last
events in the catalog, thus giving
an apparent Omori exponent $p=3$. As we know by experience that the
Omori exponent is smaller
than $2$, an exponent $p=3$ will thus be interpreted as spurious
and due to the limited duration of the catalog. This
asymptotic property, which is very different from the Omori law we 
are studying,
is a desirable property of the wavelet kernel (\ref{mngnmlq}) which provides
a clear signal of a possible problem in the data analysis. For 
example, if no event
is present in the catalog between times $t_1$ and $t_2$, and provided 
that $t_2$ is
sufficiently larger than $t_1$, then we will measure a spurious $p=3$
for time scales $a$ larger than $t_1$ and lower than $t_2$. This may lead to
some bias in the determination of $p$, that we shall come back to in 
the discussion.

\subsection{Synthetic tests}

The determination of the exponents $p$ and $H$ is performed by two algorithms
described in Appendix B. This first (respectively second) algorithm uses
the $1/H$-scaling (respectively $H$-scaling law) normalized curves.

We have performed tests of the wavelet method on synthetic catalogs generated
using the ETAS model and various modifications thereof, with and without
genuine diffusion and with or without background seismicity.
A particular result is that the wavelet method
works better the larger the background noise, up to the point
where the meaningful signal would disappear. While for large
catalogs with a significant background component, the wavelet method
is superior to the windowing method by providing more precise values 
of $p$ and $H$,
it turns out to be inferior to the windowing method for
synthetic catalogs without background when catalogs have
limited sizes, as it exhibits significant
scatter in the determination of the exponents $p$ and $H$.
Technically, this scatters
occurs due to the existence of large oscillations in the
dependence of the wavelet coefficients as a function of $R$ and/or $a$,
which makes difficult the determination of the relevant scaling intervals.
This paradoxical result reflects the larger sensitivity of the wavelet method
to the size of catalogs due to its intrinsic two-dimensional nature,
while the windowing method is more robust for small catalogs due to its
one-dimensional structure.

Comparing the $H$-method with the $1/H$-method, we note that,
due to the fact that the $H$-method uses original
curves and not their fit approximations (and are thus more subjected 
to fluctuations), the ``variance landscapes'' defined in Appendix B exhibit
much more elongated valleys around minima than with the $1/H$-method.
As will be shown below, both methods most often yield the same
results, but the $1/H$-method yields better-defined minima and should
be preferred in general.

\subsection{Results}

Table \ref{tab2wavelet} summarizes the results on the values
of $p$ and $H$ obtained from the wavelet analysis applied to 21 large
earthquakes in California. These events are the same as those studied
with the windowing method and reported in Table \ref{tabdifobs}.
Table \ref{tab2wavelet} shows that results obtained with the two
methods based on the $H$-scaling law and $1/H$-scaling law compare rather
well, except for four cases: Westmorland, Round Valley, North Palm
Springs, Mammoth Lakes. However, for each earthquake,  a detailed analysis
of the corresponding contour lines of the average variance of the collapse of
wavelet coefficients (see Figure \ref{roundvalley} for the Round 
Valley mainshock)
shows that the
$1/H$-scaling law method is in general more reliable with a more constrained
and better defined minimum. We thus tend to believe the
$p$ and $H$ values given by the $1/H$-method as more reliable.

\begin{figure}[t]
\begin{center}
\psfig{file=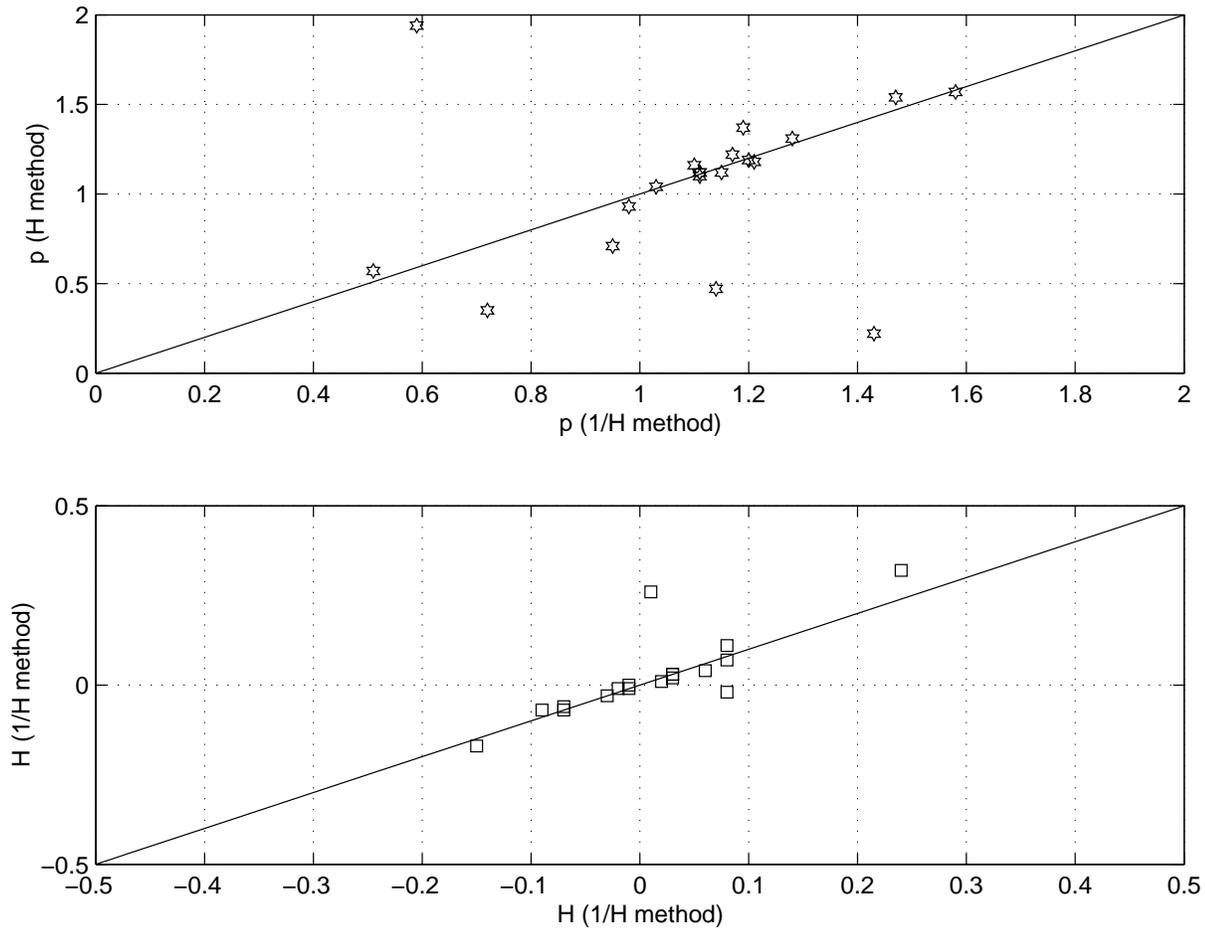,width=16cm}
\caption{ \label{Corr1}
Upper panel: correlation between the Omori exponent $p$ obtained with
the $1/H$ method (horizontal axis) and with the $H$ method (vertical
axis). The line of slope $1$ is drawn as a reference.
Lower panel: correlation between the exponents $H$ obtained with the
$1/H$ method (horizontal axis) and with the $H$ method (vertical
axis). The line of slope $1$ is drawn as a reference.}
\end{center}
\end{figure}

Figure \ref{Corr1} shows the correlation between the
exponents $p$ and $H$ obtained for each shock.
There is a good consistency between both methods.
Strictly speaking, negative values of $H$ are associated with
``anti-diffusion'', that is,
migration of aftershocks towards the mainshock. One possible reason for
our finding of negative $H$'s
is a spurious bias due to the mismatch between the mainshock epicenter
and the barycenter of the aftershock cloud.

\section{Discussion}

\subsection{General synthesis}

Our main conclusion is that, in contrast with previous claims,
diffusion of aftershocks is in general absent or very weak, at the borderline
of detection. This conclusion is reached notwithstanding our
significant efforts to develop two independent techniques which have
been optimized for extracting a signal on diffusion in the presence of
background noise. Maybe, to be fair, we should state more correctly
that our rather negative conclusion is reached precisely because
we have made large efforts to remove spurious signals. As many tests
have shown, some of which presented in section \ref{secmetobsdif},
it is easy to construct a diffusion signal in the form of a power law of
the characteristic size of the aftershock cloud as a function of time.
Our present work has stressed the importance of not jumping to
conclusions and that serious tests should be performed to assess
the reliability of a putative diffusive power law.
The simplest explanation is often that such a power law is due to
cross-over effects in the presence of inhomogeneity of the catalogs,
of their limited sizes and of the contagion induced by background and
uncorrelated seismicity. We also note that most sub-diffusion exponents
$H$ reported in previous studies as well as in the present one
are very small, in the range $0.05-0.2$.
For instance, a value $H=0.1$ implies that ten decades in time scales
are needed for each decade in space scale. This ``small exponent curse''
is one of the many explanations for the difficulty in obtaining a
clear-cut diffusion signal.

\begin{figure}[ht]
\psfig{file=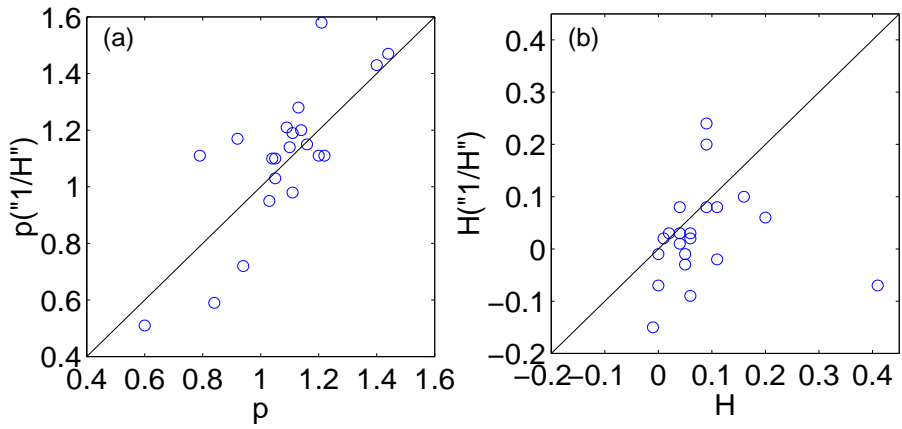,width=12cm}
\caption{\label{p1p2H1H2}
Comparison of the two methods of estimation of $p$ and $H$.
(a) Omori exponent measured by the wavelet $1/H$ method (Table 2) versus
the Omori exponent estimated by the windowing method (Table 1).
(b) Diffusion exponent measured by the wavelet $1/H$ method (Table 2) 
versus the diffusion exponent $H_r$  estimated by the windowing 
method (Table 1).}
\end{figure}

Figure \ref{p1p2H1H2} compares the values of $p$ and $H$ obtained by 
the two methods given in Tables 1 and 2. 
The results for $H$ are essentially uncorrelated between the two methods.
Considering they belong to the same aftershock sequences,
this supports the
conclusion that the diffusion coefficients are not on the whole
statistically significant.

Having said that, Tables 1 and 2 show that a few aftershock sequences
exhibit a significant and unambiguous diffusion. In order to aggregate
the information derived from the
windowing and wavelet methods, we compare the exponent $H_b$ for the former
and the exponent $H$ obtained from the $1/H$-scaling method for the latter
and qualify the existence of diffusion
(somewhat arbitrarily) when the two criteria $H_b \geq 0.1$ and 
$H_{1/H}\geq 0.05$
are simultaneously verified.
Comparing these two methods, six clear-cut cases emerge:
Westmorland ($H_b=0.12$, $H_{1/H}=0.10$), Morgan-Hill
($H_b=0.44$, $H_{1/H}=0.08$), Round-Valley
($H_b=0.11$, $H_{1/H}=0.24$), Superstition-Hill ($H_b=0.10$, $H_{1/H}=0.06$),
Joshua Tree ($H_b=0.27$, $H_{1/H}=0.08$) and Mammoth Lakes
($H_b=0.16$, $H_{1/H}=0.20$). For the other sequences, either
both windowing and wavelet analyses give a very small $H$ or they
strongly disagree with each other. The reason for the
disagreement between the two methods when it occurs is not obvious to us.
With a single method, one can quantify its systematic errors. Using
two distinct methods and comparing them enables us to quantify the 
uncertainties resulting from causes that are difficult to assess 
a priori. This is our main justification for developing two distinct
methods and for comparing their results. We note that such a strategy of
using several models emphasizing different physical
mechanisms with distinct implementations is well-known and largely used in meteorological 
forecasts, precisely with the aim of accounting for the unknown
or non-understood sources of uncertainties. 

It may be helpful to put these results in the light
offered by the  ETAS model [{\it Helmstetter and Sornette}, 2002b] whose main
predictions are summarized in Appendix A. The most robust prediction
is that one should expect aftershock diffusion
when the Omori's exponent $p$ is less than $1$, because this value
signals the existence of a cascade of triggering which is the
mechanism at the origin of diffusion in the ETAS model.
In the above list of $6$ clear-cut cases, three (Morgan-Hill,
Round-Valley and Mammoth Lakes) have an exponent $p$ smaller than $1$
according to both methods, while the three others have a $p$-value
larger than or very close to $1$. This fifty-fifty deadlock seems to
discredit the usefulness of the ETAS model for this problem. It may
be useful to look in more detail at the results of each
method separately.

\subsection{Discussion of the results in the
light of the ETAS model}

Most sequences shown in Table 1 for the windowing method
and in Table 2 for the wavelet method are in the regime
$p>1$ and are characterized by $H \approx 0$. 
These results are compatible
with the predictions of the ETAS model that no diffusion
should be observed if the Omori exponent is larger than $1$.
Indeed, a measured value $p>1$
can be interpreted as belonging to the sub-critical regime $\nu<1$
(where $\nu$ is the average number of events triggered per triggering event)
such that the characteristic time $t^*$ defined by (\ref{tjmjs}) is small 
and the relevant time window covers mostly the regime $t>t^*$ for
which the cascade process is exhausted and diffusion is absent
(Appendix A).

As seen in Table 1 for the windowing method and in Table 2 for the
wavelet method, a few aftershock sequences are characterized by a 
small $p<1$ exponent. As we have said, according to the ETAS model
(see Appendix A), this is the relevant regime for
observing aftershock diffusion.

\begin{figure}
\begin{center}
\psfig{file=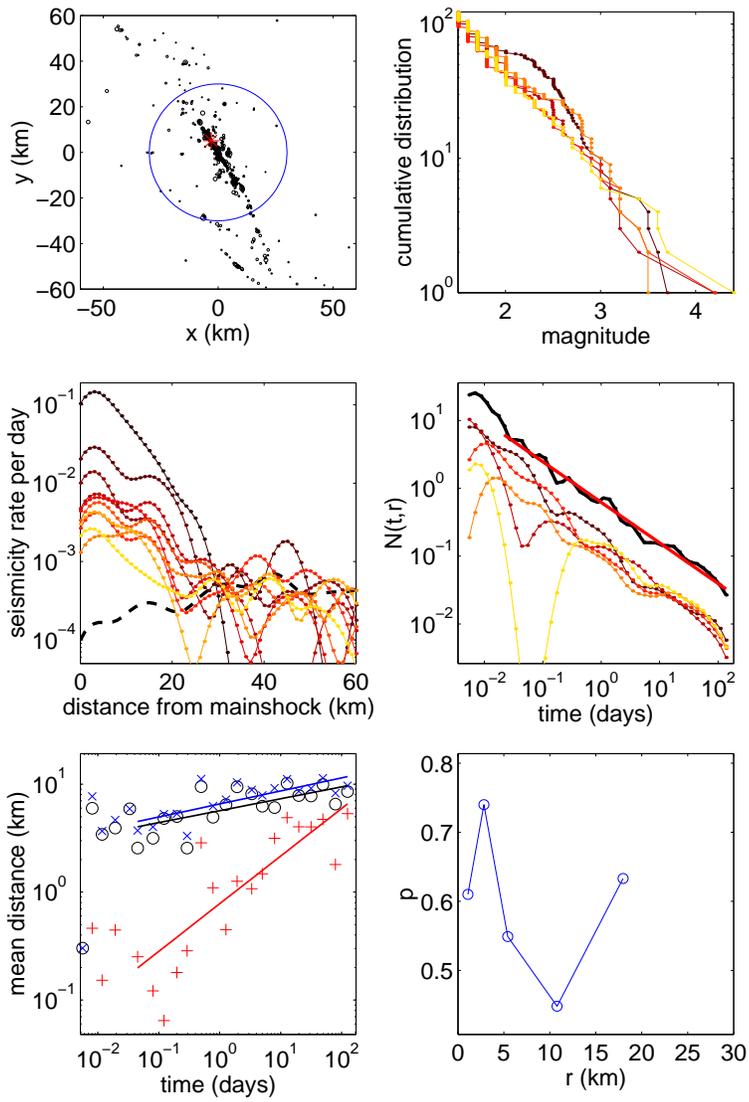,width=10cm}
\caption{ \label{MHd} Analysis of the Morgan-Hill aftershock sequence.
Same legend as in Figure \ref{landersdif}.}
\end{center}
\end{figure}

\begin{figure}
\begin{center}
\psfig{file=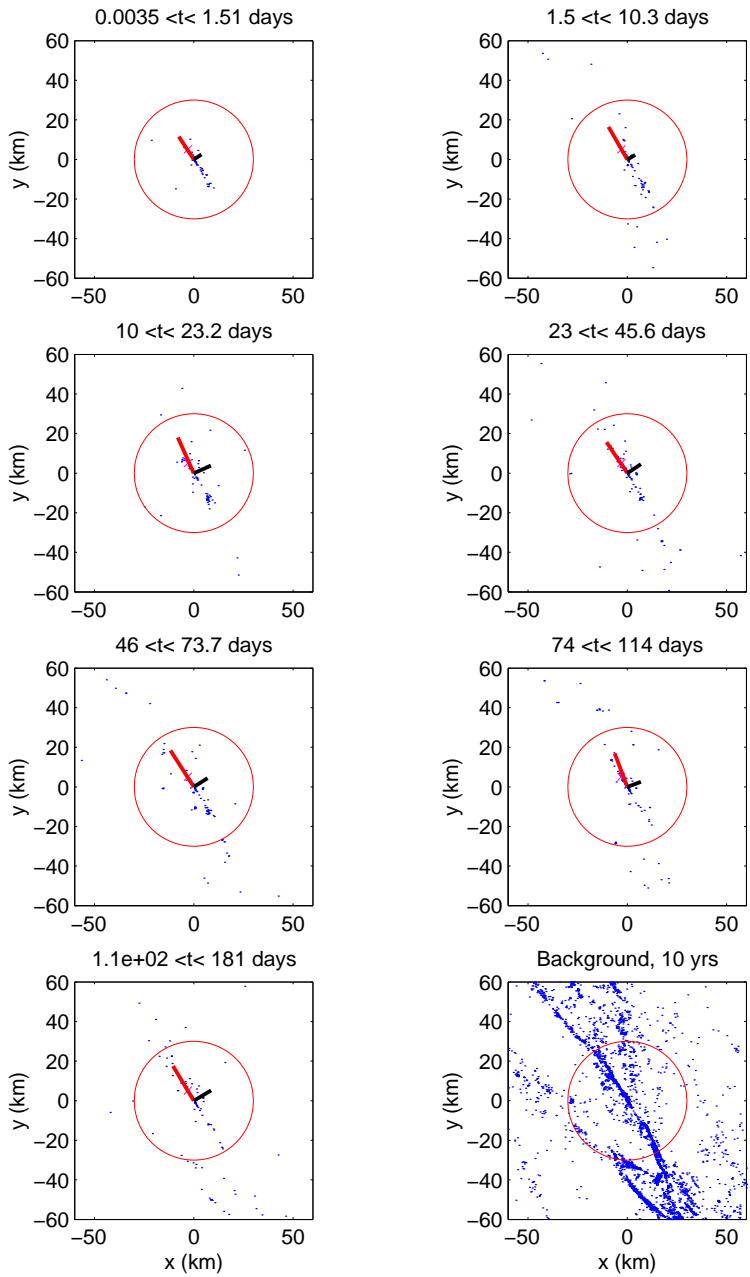,width=10cm}
\caption{ \label{MHm} Map of the Morgan-Hill aftershock sequence.
Same legend as in Figure \ref{landersmap}.}
\end{center}
\end{figure}

The Morgan-Hill sequence, analyzed in Figures \ref{MHd} and \ref{MHm} and
which has the smallest $p$-value with both the windowing method
($p=0.6$) and the $1/H$-wavelet method ($p=0.5$), has a small but 
significant diffusion exponent measured with the windowing method.
We obtain $H_r \approx H_a \approx 0.1$
estimated by the aftershock distances from the barycenter or the
large elliptical axis, and a large value $H_b=0.44$ obtained by
using the time evolution of the small elliptical axes $b$ which
measures the average distance of aftershocks from the rupture plane.
The value of $H=0.08$ obtained by the wavelet $1/H$-method is similar.
The larger value $H_b=0.44$ could be due to the fact that the diffusion
perpendicular to the fault is less perturbed by
the aftershocks along the whole length of the fault
which occur in absence of genuine diffusion.
For this aftershock sequence, the empirically determined values of
$p=0.6$ and $H_b=0.44$ would correspond for instance
to $\theta \approx 0.4$ and $\mu \approx 1$ (see equation (\ref{Hdef}))
according to the ETAS model [{\it Helmstetter and Sornette}, 2002b],
where $\theta$ and $\mu$ are defined in equation
(\ref{nmgjedl}) in Appendix A.

\begin{figure}
\begin{center}
\psfig{file=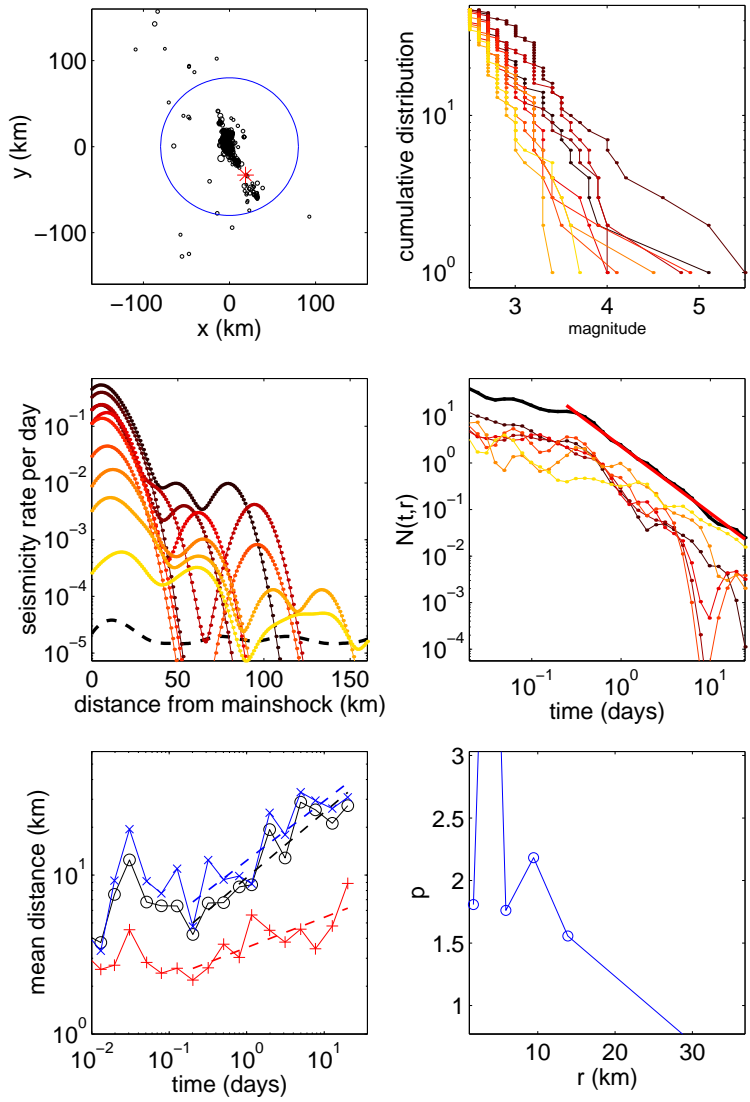,width=10cm}
\caption{ \label{IVd} Analysis of the Imperial Valley aftershock sequence.
Same legend as in Figure \ref{landersdif}.}
\end{center}
\end{figure}

\begin{figure}
\begin{center}
\psfig{file=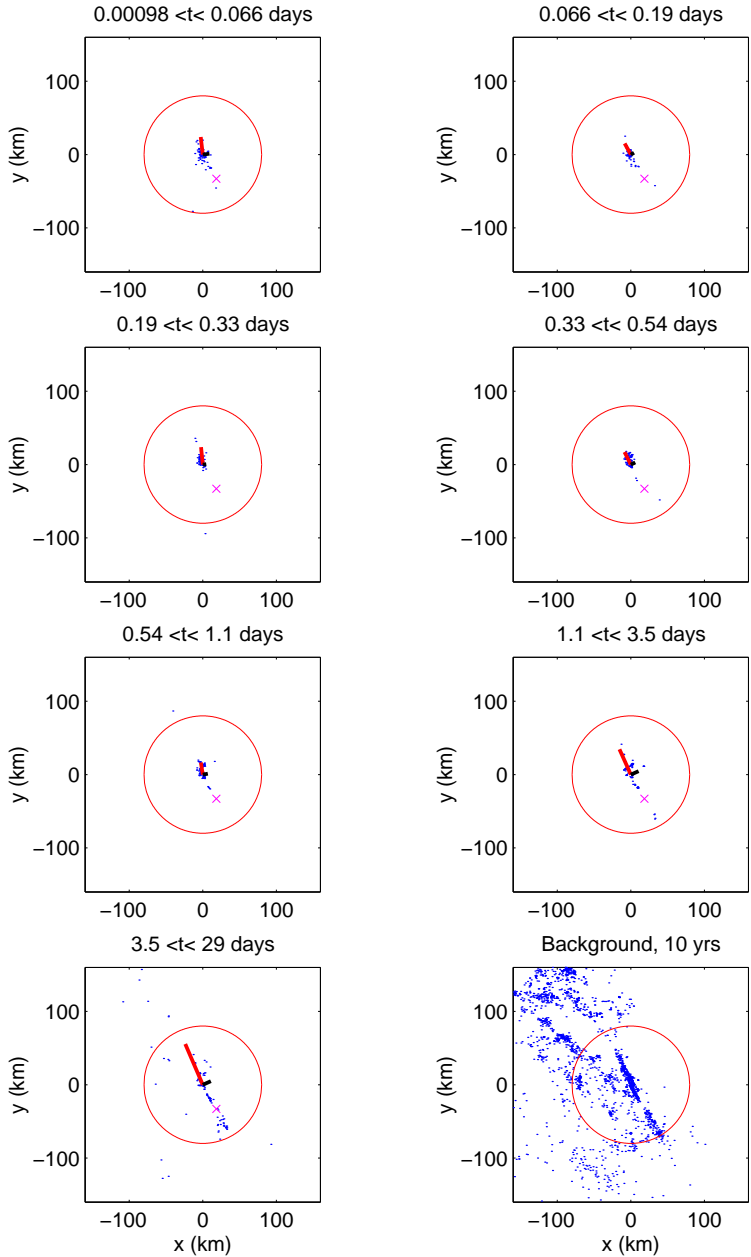,width=10cm}
\caption{ \label{IVm} Same as Figure \ref{landersmap}
for the Imperial Valley aftershock sequence. Note that the epicenter
shown as a cross is far off from the locations where aftershocks cluster.
This justifies our use of the aftershock barycenter as a more natural
reference point for measuring diffusion.}
\end{center}
\end{figure}

The other sequences in Table 1 with $p<1$ for which the cascade model predicts
the existence of diffusion are Kern-County, Round-Valley, Oceanside and
Mammoth Lakes (see table \ref{tabdifobs} for the corresponding
parameters). Except for Oceanside,
the corresponding $p$ and $H_b$ values are compatible with the
prediction of the ETAS model [{\it Helmstetter and Sornette}, 2002b] 
with $\mu \approx 1$ leading to $H_b \approx 1-p$.
While these results are suggestive for the validity
of the ETAS predictions, more disturbing is the fact that
large values of the diffusion exponents $H$ are found for $p>1$
(see  Table \ref{tabdifobs}).
The Imperial Valley sequence is a case in point, with the largest $p=1.44$
and large diffusion exponents $H_r=0.41, H_a=0.37$ and $H_b=0.19$.
Its detailed analysis is shown in Figures \ref{IVd} and \ref{IVm}.
For this sequence, one can clearly observe an expansion of the aftershock
area when comparing the distance distribution at different times. There is
also a clear decrease of the exponent $p$ with $r$ as predicted by the ETAS
model as a signature of the aftershock cascade leading to diffusion, but
this should be associated with $p<1$ and not with $p>1$ as found 
here. It is true that a significant diffusion exponent $H>0$ with
$p>1$ can be observed in the ETAS model in the crossover regime for 
$t \approx t^*$ where $p$ is already larger than 1, but where a diffusion 
of seismic activity is still observed. Indeed, synthetic aftershock sequences
 in the sub-critical regime exhibit a diffusion of seismic activity which 
persists up to $t\approx 100~ t^*$ even if the Omori exponent in larger $1$.
But the diffusion exponent $H$ in the crossover regime for $t \approx 
t^*$ should be smaller than in the early time $t < t^*$ regime when $p$ is 
smaller than $1$. We thus do not fully understand the origin of this discrepancy.

\begin{figure}
\begin{center}
\psfig{file=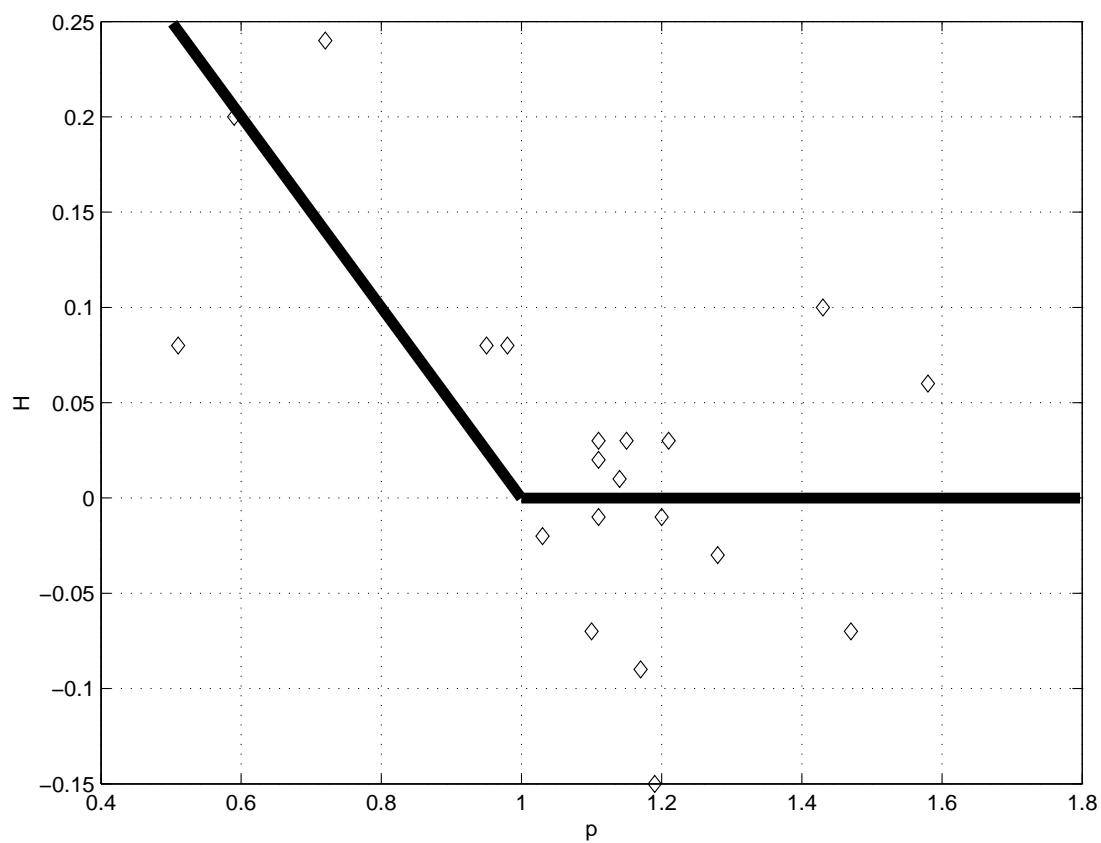,width=16cm}
\caption{\label{Corr2}
Diffusion exponent $H$ as a function of the Omori law exponent $p$
obtained with the $1/H$ method for all the aftershock sequences
described in table \ref{tab2wavelet}. The thick lines are the
approximative predictions of the ETAS model for $\mu \geq 2$
(\ref{Hdef}), assuming that $p=1-\theta$ if $p<1$ and $p=1+\theta$
if $p>1$.}

\end{center}
\end{figure}

Using the wavelet method, only five sequences are in the regime $p<1$ 
where diffusion is predicted to occur according to the ETAS model and, 
there, the expected correlation is weak and noisy.
The evidence of aftershock diffusion is very weak as
$75\%$ of the aftershock sequences seem to be in the non-critical
regime ($t>t^*$ and $p>1$) characterized by an absence of diffusion.
The remaining five sequences are loosely compatible with
the existence of diffusion and the quantitative values are
consistent with the predictions of [{\it Helmstetter and Sornette},  2002b], 
when taken collectively. Figure \ref{Corr2} tests the possible 
existence of a correlation between the
diffusion exponent $H$ and the Omori law exponent $p$
obtained with the $1/H$ method for all the aftershock sequences
described in table \ref{tab2wavelet}. The thick lines are the
prediction of the ETAS model [{\it Helmstetter and Sornette}, 2002b] that
$H \approx0$ for $p>1$ and $H=(1-p)/2$ for $p<1$, obtained by choosing $\mu
\geq 2$ and by assuming $p=1-\theta$ (which is correct if $t \ll t^*$).
In the five cases with $p<1$, we find positive exponents $H$ in the
range from $0.08$ to $0.24$. A linear fit over these five events with
$p<1$, forced to pass through the ``origin'' $(p=1, H=0)$ gives
$H=(p-1)/2.52$, which must be compared with the prediction
$H=(1-p)/2$.
In the other 15 cases with $p>1$, $H$ is very noisy with almost
as many negative and positive values in the range
from $-0.15$ to $0.1$, and almost vanishing average.

In conclusion, the results obtained with the two methods
do not show any significant correlation between $H$ and $p$, in
contradiction with the prediction of the ETAS model.
This disagreement between the observations and the theory
may result from the insufficient number of aftershocks
available.
The small number of events used yields a large uncertainty 
of the measure of $H$ and $p$, as shown by the large difference in
the value of $p$ and $H$ obtained for some sequences 
using the different methods (windowing, $H$ and $1/H$ wavelet method).
In addition, these are other factors, such as the fact
that the typical mainshock-aftershock distance increases with the 
mainshock magnitude, or the geometry of the rupture, 
which are taken into account neither in the ETAS model nor 
in the empirical analysis,
but which may significantly alter the results.

\section{Factors limiting the observation of diffusion \label{geom}}

This section examines important limitations of both the theoretical
analysis of {\it Helmstetter and Sornette} [2002b] and our present
 study of California seismicity and discusses possible remedies.

\subsection{Independence between the mainshock size and the
aftershock cluster size and selection rules \label{gemoakmab}}
 
A limitation of the analytical approach developed
in  [{\it Helmstetter and Sornette}, 2002b] is that
the distribution of distances between a mainshock and its aftershocks
are assumed independent of the mainshock magnitude.
However, it is a well established property of aftershock sequences
that the size of the aftershock area is  approximately proportional
to the mainshock rupture length [{\it Utsu}, 1961; {\it Kagan}, 2002].

We can modify the ETAS model to include a dependence
between the mainshock magnitude and the aftershock size,
as observed in real seismicity, in order  to take into account
the extended rupture length of the mainshock.
In this goal, we modify the distance distribution
$\sim 1/(r+d)^{1+\mu}$ defined in (\ref{nmgjedl}) of Appendix A
by taking the characteristic size $d(M)$ proportional to
the mainshock rupture length $d(M) \propto L \sim 10^{0.5 M}$.
This means that an aftershock of generation $i$ is created
by an aftershock of generation $i-1$ of magnitude $M_{i-1}$
at a distance $r$ from it via the rate (\ref{nmgjedl}) with the 
distance distribution $\sim 1/(r+d(M_{i-1}))^{1+\mu}$.

We can understand intuitively the effect of introducing a dependence
between $d$ and the magnitude, knowing the solution $N(t,r)$
given by {\it Helmstetter and Sornette} [2002b] for a constant $d$-value.
Starting from a mainshock
of magnitude $M$ at the origin, the typical distance between aftershocks
of the first generation and the mainshock is $d(M) \propto L$, where
$L$ is the rupture length of the mainshock. Aftershocks of second and
later generation have a smaller magnitude on average, otherwise they
do not qualify as aftershocks of the first event, by our present definition.
We can thus define an average rupture size $\langle \ell \rangle < L$ for
the aftershocks. As long as the average size of the aftershock zone
$R(t)$ is smaller than $L$, the aftershocks of the first
generation dominate the spread of seismic activity following the mainshock
and diffusion is not observable.
At large times, when the aftershock area of second and
later generations becomes larger than the influence
zone of the mainshock of size $L$ and when most aftershocks 
are triggered indirectly by aftershocks of the mainshock, the 
diffusion for the coupled model should be the same as for the 
decoupled model studied in  [{\it Helmstetter and Sornette}, 2002b]. 
The dependence of $d$ with the progenitor's magnitude thus introduces a
crossover time $\tau$ such that there is no significant diffusion
below $\tau$ because the aftershock spatial expansion is dominated
by aftershocks of the first generation while for time larger than $\tau$
the multiple cascades of aftershocks dominates the spatial excursion of
aftershocks and we recover the diffusion law (\ref{Rt}) of the decoupled
model. This gives the equation for $\tau$:
\be
\langle \ell \rangle \Bigl({\tau \over c} \Bigr)^H \approx L~,
\ee
leading to
\be
\tau = c \left( {L \over \langle \ell \rangle} \right)^{1/H}
\ee
where $L$ is the mainshock size and $\langle \ell \rangle$ is the 
average rupture length of aftershocks. For large mainshocks 
$L \gg \langle \ell \rangle$ and slow diffusion, this characteristic
time $\tau$ can be very large and, in practice, diffusion may be 
unobserved in real aftershock sequences.

\subsection{Rules of aftershock selection}

\begin{figure}[ht]
\psfig{file=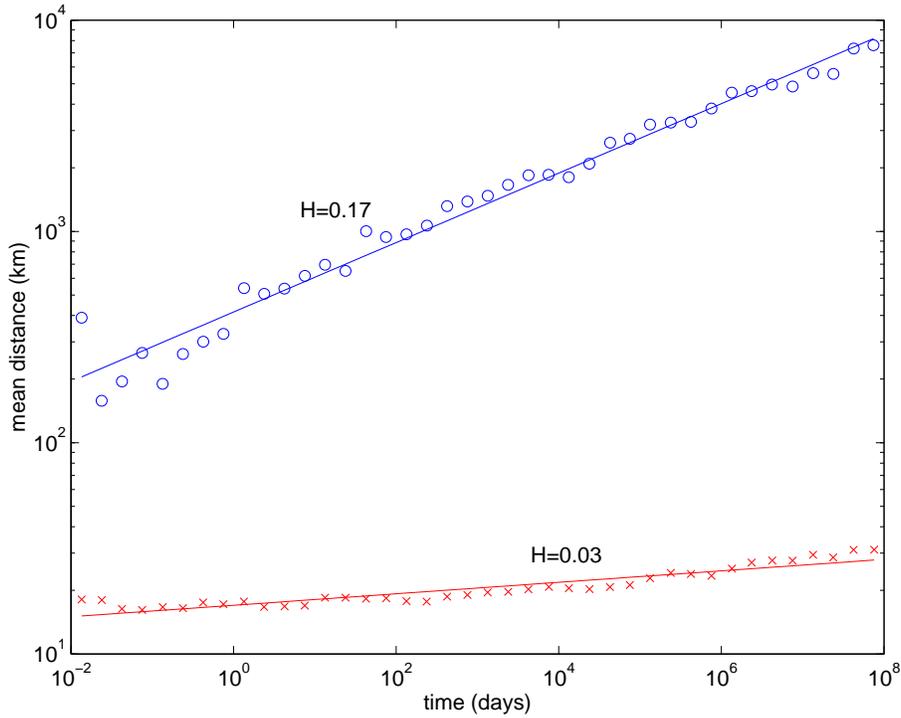,width=12cm}
\caption{\label{HdL}Average distance $R(t)$ between a mainshock of 
magnitude $M=6$ and its aftershocks for numerical simulations of the
ETAS model with $n=1$, $\alpha=0.5$, $b=1$, $c=0.001$, $\theta=0.2$,
$m_0=0$ and  $\mu=1$, obtained by averaging over 1000 simulations.
The circles show the results for $d=10$ km, independently of the 
mainshock size, and without any constrain on aftershock selection. 
The diffusion exponent $H=0.17$ is close to the prediction
$H=\theta/\mu=0.2$. The crosses correspond to another simulations 
of the ETAS model with the same parameters except that the
characteristic distance $d$ now depends on the magnitude
of each event according to $d=0.01 \time 10^{0.5 M}$. For this 
simulation, we have also selected aftershocks up to a distance 
of 100 km (10 times the mainshock rupture length) and we have 
rejected all aftershock sequences containing at least an event 
larger than the mainshock. All these factors weaken
the diffusion by comparison the the prediction of the ETAS model.}
\end{figure}

There are two other factors that may further weaken the observed diffusion
in comparison
with the prediction of the ETAS model given in [{\it Helmstetter and Sornette},
2002b]. First, the constraint that the aftershocks must be smaller than the
mainshock induces an under-estimation of the true diffusion exponent.
If we include this rule of aftershock selection in the numerical
simulations,
the average seismicity rate is smaller than the predicted seismicity rate
and decreases as $1/t^{1+\theta}$ at large times instead of the
theoretical prediction $ N(t) \sim  1/t^{1-\theta}$ for $t<t^*$ or $n=1$. 
This effect induces a slower diffusion at large times when the observed 
Omori exponent becomes larger than 1. The other factor which may lower 
the diffusion in the real aftershock sequence
is the introduction of a  maximum distance $r_{max}$ for the selection
of aftershocks close to the mainshock. Introducing this rule in the
simulations of the ETAS model obviously lowers the diffusion exponent 
by comparison with the true
exponent, because it rejects all aftershocks triggered at large distance
from  the mainshock which have an important role in the diffusion process,
especially for small value of $\mu<2$.
The effect of the rupture geometry and the rules of aftershock selection
are illustrated in Figure \ref{HdL} for numerical simulations
of the ETAS model obtained for $\alpha=0.5$, $b=1$, $n=1$, $\mu=1$,
$c=0.001$ day, $m_0=0$, $M=6$, $d=0.01 \times 10^{0.5 M}$ km.
This Figure shows that the combined effect of the dependence between
the influence distance $d$ and the rupture length $L$, and the rules
of aftershock selection results in a strong reduction of the diffusion
in comparison with the predictions given in [{\it Helmstetter and Sornette},
2002b].

\subsection{Influence of the fault geometry \label{gemoakma}}

Another important limitation of the ETAS model and of other models
of aftershocks such as the rate-and-state friction model 
[{\it Dieterich}, 1994] is that these models do not take into
account the anisotropy of the spatial distribution of aftershocks
nor the localization of earthquakes on a fractal
fault network. These factors may also have an important influence 
on aftershock diffusion.

There is strong evidence that earthquakes occur on faults
and faults are organized in a complex hierarchical network 
[{\it Ouillon et al.}, 1996].
A simplified representation of this network uses fractal geometry.
A better description of aftershock diffusion, if any, should thus
deal with the problem of diffusion on a fractal network. Some general
results have been obtained in the physical literature on this problem
of diffusion of probe particles in non-Euclidean fractal spaces
(see {\it Bouchaud and Georges} [1990], {\it Sahimi} [1993; 1994]
and references therein). The main
consequence is that the diffusion exponent $H$ may
be modified to take into account the fractal geometry
through the introduction of the so-called spectral dimension (which is
often different from the geometrical fractal dimension). 
In general, this leads to reduced diffusion when measured with
Euclidean distances. Another possible caveat is the anisotropy of 
fault networks, resulting from the localization properties
of mechanical systems which tend to organize oriented shear bands.
For instance, there is a very strong South-East to North-West preferred
orientation of the San Andreas fault network in California. 

While previous empirical
analyses including the present one as well as the ETAS model of cascades
of triggering have neglected geometry, our tests suggest that geometry
is a crucial ingredient. By geometry, one should include both
the effect of a possible
localization of a fraction of aftershocks on mainshock rupture faults
and their localization of pre-existing hierarchical fault networks,
as well as the dependence of the range over which aftershocks are triggered
by the mainshock on the mainshock magnitude.

\section{Conclusion}

We have analyzed 21 aftershock sequences of California and found that the
diffusion of seismic activity is very weak, when compared with 
previous studies.
For most sequences, the spatial distribution of aftershocks is mostly
limited to the mainshock rupture area. The rate of aftershocks is very
small at distances larger than the rupture length, even at large times
after the mainshock.

In the introduction, we noted that aftershock diffusion is predicted by any
model that assumes that the time to failure increases if the applied
stress decreases. Our conclusion that aftershock diffusion is weak at best
or non-existent suggests that the physical process controlling the time of
failure depends weakly on the magnitude of the stress change in the regime
where this stress change is sufficient to trigger new events.

We have examined the hypothesis that aftershock diffusion may result from
multiple triggering of secondary aftershocks. In principle, this
mechanism offers clear quantitative predictions, if one controls
the parameter regime of the model. One problem is that
our theoretical and numerical studies of
cascade processes indicate that most predictions are very sensitive 
to small changes in the parameters of the seismic activity which
cannot be easily determined from seismicity catalogs.
This variability of the ETAS parameters from sequence to sequence may 
thus rationalize the variability of the diffusion exponent from one 
sequence to another one. In conclusion, the large uncertainty on the 
estimation of the Omori exponent $p$ and the diffusion exponent
$H$, resulting from different factors (small number of events, small 
available time and space scale, background seismicity and fault geometry, 
high fluctuations from one sequence to another one), does not allow
us to conclude clearly the validity or 
the irrelevance of the mechanism of triggering 
cascade embodied by the ETAS model in describing aftershock diffusion.

\acknowledgments
We thank D. Marsan for providing his code and the preprints of his papers
and for several exchanges and J.-R. Grasso for useful discussions.
This work is partially supported by NSF-EAR02-30429, by
the Southern California Earthquake Center (SCEC) and by
the James S. Mc Donnell Foundation
21st century scientist award/studying complex system. We acknowledge
SCEC for the earthquake catalogs.

\section*{Appendix A: Summary of main results of the ETAS model
concerning aftershock diffusion}

The ETAS (Epidemic-Type Aftershock Sequence) model was introduced
in {\it Ogata} [1988] and in {\it Kagan and Knopoff} [1981; 1987] 
(in a slightly different form) based on mutually exciting point processes
introduced by Hawkes [{\it Hawkes}, 1971; 1972; 
{\it Hawkes and Adamopoulos}, 1973]. Contrary to
what its name may imply, it is not only a model of aftershocks
but a general model of triggered seismicity.

This parsimonious physical model of multiple cascades of
earthquake triggering avoids the division between foreshocks,
mainshocks and aftershocks because it uses the same laws to describe all
earthquakes without distinction between  foreshocks,
mainshocks and aftershocks. It is found
[{\it Helmstetter and Sornette}, 2002a,b,c; 2003a;
{\it Helmstetter et al.}, 2003] 
to account surprisingly well for (i) the
increase of rate of foreshocks before mainshocks (ii) at large distances
and (iii) up to decades before mainshocks, (iv) a change of the
Gutenberg-Richter law from a concave to a convex shape for foreshocks,
and (v) the migration of foreshocks toward mainshocks. The
emerging concept is that the cascade of secondary, tertiary and
higher-level triggered events gives rise naturally to long-range and
long-time interactions, without the need for any new physics. This
emphasis on cascades of triggered seismicity provides a general
understanding of the space-time organization of seismicity and offers
new improved methods for earthquake prediction.

In the ETAS model, a main event of magnitude $m$ triggers its own
primary aftershocks according to the following distribution in time and space
\begin{equation}
\phi_m (r,t)~ dr~ dt = K~ 10^{\alpha m}
~\frac{\theta~c^{\theta}~dt}{(t+c)^{1+\theta}} ~
\frac{\mu~d^{\mu}~dr}{(r+d)^{1+\mu}}~,
\label{nmgjedl}
\end{equation}
where $r$ is the spatial distance to the main event (considered as a
point process). The spatial regularization distance $d$ accounts for
the finite rupture size. The power law kernel in space with
exponent $\mu$ quantifies the fact that
the distribution of distances between pairs of events is
well described by a power-law [{\it Kagan and Jackson}, 1998]. 
In addition, the magnitude
of these primary aftershocks is assumed to be distributed according
to the Gutenberg-Richter law with slope $b$. The ETAS model assumes that
each primary aftershock may trigger its own
aftershocks (secondary events) according to the same law, the secondary
aftershocks themselves may trigger tertiary aftershocks and so on,
creating a cascade process. The exponent $1+\theta$ is not the observable
Omori exponent $p$ but defines the local (or direct) Omori law.
The whole series of aftershocks, integrated over the whole space,
can be shown to lead to a ``renormalized'' (or dressed)
Omori law, which is the total observable Omori law [{\it Helmstetter 
and Sornette}, 2002a].
This global law is different from the direct Omori law
$1/(t+c)^{1+\theta}$ in (\ref{nmgjedl}):
the observable (dressed) Omori exponent crosses over smoothly from the
value $1-\theta$ below a characteristic time $t^*$ to
$1+\theta$ at times much larger than $t^*$
\be
t^* = c ~\left({\nu~\Gamma(1-\theta) \over |1-\nu|}\right)^{1/\theta}
\label{tjmjs}
\ee
where the branching ratio $\nu$, defined as the average number
of aftershocks per earthquakes, is a function of the parameters of
the ETAS model $\nu=K~b /( b-\alpha)$.
The renormalization of the direct Omori law with exponent $1+\theta$
into the observable dressed Omori law with exponent $1-\theta$ (for 
$t\leq t^*$)
results from the cascade process. Intuitively, it is clear that
the existence of cascades of secondary aftershocks may lead to
observable diffusion, analogously to random walks whose succession of
jumps create diffusion upon averaging. This analogy was
established in [{\it Helmstetter and Sornette}, 2002b] 
which predicted different diffusion
regimes according to the values of the model parameters.
The simplest mathematical characterization of diffusion is through
the evolution of the characteristic size $R$ of the aftershock cloud
as a function of time $t$ since the main shock:
\be
R \sim t^H~,
\label{hakllw}
\ee
where $H$ is the diffusion exponent (equal to $1/2$ for classical diffusion).
The theory and numerical
simulations developed in [{\it Helmstetter and Sornette},  2002b]
predict, for $t < t^*$ and $\theta <1$ :
\ba
H &=& \theta/2 ~~~ \mbox{for} ~~\mu>2 \\ \nonumber
H &=& \theta/\mu ~~~ \mbox{for}~~  \mu <2 ~.
\label{Hdef}
\ea
where
\be
r^*=\sigma \left({\nu \over 1-\nu}\right)^{1/\mu}~,
\label{gnjgrkd}
\ee
and $\sigma$ is proportional to the
spatial regularization distance $d$ in (\ref{nmgjedl}) up to a numerical
constant function of $\mu$ [{\it Helmstetter and Sornette}, 2002b].
In all cases, the diffusion saturates progressively as $t$
becomes much larger than $t^*$.  Here, the important message is that,
despite the fact that time and space are
uncoupled in the direct triggering process (\ref{nmgjedl}),
the succession of cascades of events can lead to a macroscopic
coupling of time and space, that is, to diffusion.
This is illustrated in Figure \ref{map} which presents
results from numerical simulations of the ETAS model to show
how cascades of multiple triggering can produce aftershock diffusion.
Since the condition $t<t^*$ also
ensures that the observable Omori exponent $1-\theta$ is different from
the direct exponent $1+\theta$, this triggering cascade theory predicts that
diffusion should be observed most clearly when Omori's exponent $p$ is
smaller than $1$. Note that, for $p$ close to $1$ as
often observed empirically, that is for $\theta$ small,
the predicted diffusion exponent $H$ is significantly smaller than $1/2$.
The results in [{\it Helmstetter and Sornette}, 2002b] were
obtained for a one dimensional process, but most results,
in particular the diffusion law (\ref{hakllw}),
are valid in any dimension. Another complication elaborated upon below
is the fractal geometry of fault networks.

A possible caveat in the predictions of the ETAS model given in 
[{\it Helmstetter and Sornette}, 2002b]  is that
we predict only the average behavior of the space-time
distribution of seismicity. In the regime where $\alpha/b>0.5$
relevant to real seismicity [{\it Helmstetter}, 2003], we observe huge 
fluctuations of the
seismicity rate around the average  that may weaken the usefulness 
of predictions based on ensemble averages. Having said that, we have 
verified by intensive numerical simulations that the prediction 
$H \geq (1-p)/2$ remains valid in all cases including the large 
deviation regime $\alpha > b/2$. Thus, for the problem treated here of
detecting a possible aftershock diffusion, this issue is not a limiting point.

Our tests on real aftershock sequences are interpreted in particular
in the light shed by the ETAS model.
We stress that the analysis of real data is much more
difficult than the study of synthetic sequences, due to the smaller number
of earthquakes available, the presence of background activity,
the effect of geometry and problems
of catalog completeness especially just after large mainshocks. In addition,
real seismicity is probably more complicated than assumed by the ETAS model.
These limitations imply that it is difficult to obtain reliable
quantitative results on the diffusion exponent. However, a few
qualitative predictions of the ETAS model should be testable in real data:
\begin{itemize}
\item only sequences in the ``early'' time regime $t<t^*$ characterized  by
an Omori exponent $p<1$ should diffuse;
\item the diffusion of seismic activity should be related to a
decrease of the
Omori exponent $p$ as the distance $r$ from the mainshock increases;
\item the characteristic size of the cluster is expected to grow according
to expression (\ref{hakllw}) with the diffusion exponent $H$ positively
correlated with the $\theta$-value.
\end{itemize}

\section*{Appendix B: Implementation of the wavelet method}

\subsection*{Method using the $1/H$-scaling law normalized curves}

The determination of the exponents $p$ and $H$ is performed by
using the following steps when using the $1/H$-scaling law normalized curves:
\begin{enumerate}
\item Compute $C_{a}(R)$ as a function of scale $a$ for a series of
given $R$ values. Typically, we considerer $a$ values varying with a
multiplicative factor of $1.1$, while $R$ varies with a multiplicative
factor of $1.01$. This last value will be justified below. For each $R$
value we thus obtain a curve showing the variation of $C_{a}(R)$ with
$a$. Those curves are called $R$-curves.

\item Select the range in scale $a$ over which all $R$-curves display 
a power-law
behavior as a function of $a$. Note that the exponent of the
power law may vary from one $R$ to another $R$ (which is the hallmark 
of an underlying
diffusion process).
We could also select a different range of $a$ for each value of $R$, 
but this would
drastically complicate the data processing.

\item For each $R$-curve, fit $C_{a}(R)$ over the selected range in $a$
by a power law. This step proves
necessary as some data sets can sometimes display strong fluctuations which
will ultimately alter
the results (this has been checked on synthetic data sets). Each
interval in $a$ for each $R$-curve is now replaced
by its power-law fit approximation.

\item Choose a trial $(p,H)$ and normalize each curve
according to these exponents and their respective $a$ and $R$ values
according to (\ref{gmjmwsw}) ($1/H$-scaling law).
The normalized time scale axis is then re-sampled using
a geometrical sampling with a multiplicative factor of $1.1$, so that
all normalized R-curves are defined at common abscissae.

\item For each normalized value of the time scale, search for all normalized
power law approximations of $R$-curves which are
defined at that value. Let us assume there are $N_s$ such segments,
each one corresponding to a different $R$-curve (thus $R$ value).

\item Compute the average value of these normalized $R$-curves
that are defined at the same normalized time scale, as well as
the associated variance. If none or only one power law approximation 
of $R$-curves
is defined for the considered normalized time scale, the calculation is
not performed as the variance cannot be estimated. Then,
go to the next normalized time scale.

\item When all values of the normalized time scale have been considered,
compute the average of all the variances that have been defined up
to now. This average variance gives us an estimation of how the
various normalized segments collapse on top of each other in the goal
of defining a single master curve. It is approximately equivalent to the
square of the average width defined by superimposing all the 
normalized curves.

\item This algorithm is implemented using a $(p,H)$ grid, with $p$
varying from $0$ to $2$ by
steps of $0.01$, while $H$ varies within $[-1:1]$, with the same
step. When $H=0$, no
computation is made. A systematic search provides the couple of
exponents $(p,H)$ with the lowest average variance, i.e., the
best collapse of the wavelet coefficients as a function
of time scale and distance. For $H=0$,
the value of the variance for any couple $(p,0)$
is estimated as the mean
of the variances obtained for $(p,-0.01)$ and $(p,0.01)$.
This estimation has no real statistical meaning but is useful
for representation purpose. The top left panel of Figure
\ref{roundvalley} constructed for the Round Valley mainshock show the
contour lines representing equivalues of the average variance
(quantified in logarithmic scale in base $10$).

\item Once the best $(p,H)$ couple has been found, the
normalized wavelet coefficients $C_{a}(R)$ are calculated as a function of $a$
using the original non-normalized real curves (and not their power-law
fit approximations). This is performed for the purpose of visualizing
the predicted collapse of the wavelet coefficients on the
real data-set, as shown in the left lower panel of Figure \ref{roundvalley}
constructed for the Round Valley mainshock.

\end{enumerate}

Note that this collapse method applied to the $1/H$-scaling law
can not work for $H=0$ strictly, since $1/H$ diverges.
To address this minor technical problem,
we choose a very small logarithmic step for the $R$
values to allow us to consider $H$ values as small as $\pm 0.01$.
Indeed, the smaller is $|H|$, the more dilated is the normalized
time scale axis. If $H$ is too small, the scaling regions of two
successive wavelet coefficients for two successive $R$ values
will undergo so much offset that they will not be defined on any
common normalized time scale, preventing an estimation of an
average variance. Considering small values of $H$ thus necessitates
the computation of wavelet coefficients for successive $R$ with a very
small step in $R$ (hence the choice of a multiplicative factor of $1.01$).

\subsection*{Method using the $H$-scaling law normalized curves}

\begin{figure}[t]
\begin{center}
\psfig{file=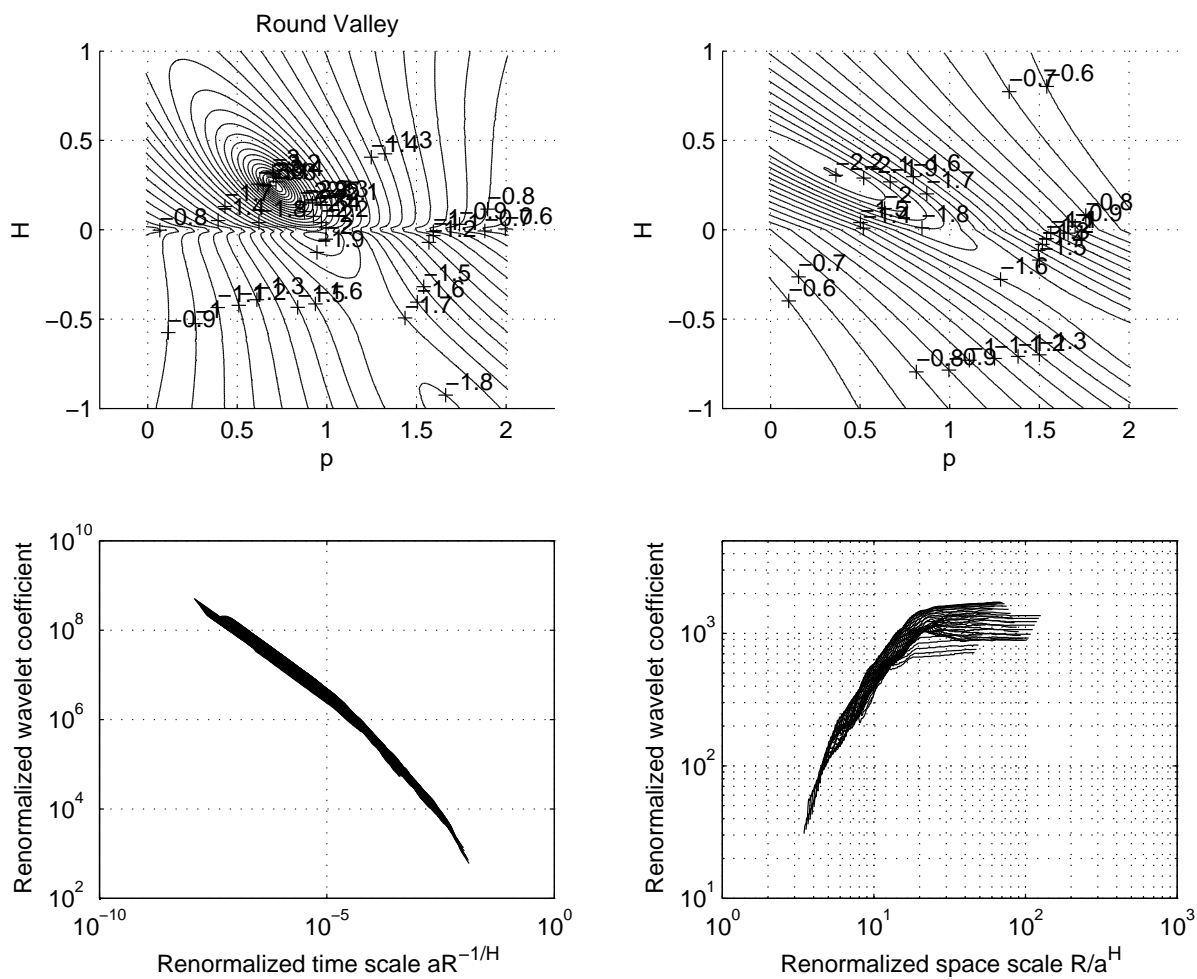,width=16cm}
\caption{ \label{roundvalley}
Determination of the exponents $H$ and $p$ for the Round Valley
mainshock using the $1/H$-scaling method (upper and lower left panels)
and the $H$-scaling method (upper and lower left panels).
The upper panels represent the contour plots in $\log_{10}$ scale
of equi-values of the average variance of the matching of wavelet
coefficients (calculated as a function of time scale for different
distances $R$; see text for details) as a function of trial values of
$p$ and $H$.  The minimum determines our best estimation for
$H$ and $p$ for this sequence. We get two estimates, one using the
$H$-scaling method ($p=0.35$, $H=0.32$) and another using the
$1/H$-scaling method ($p=0.72$, $H=0.24$). Note that the minimum
found by the $1/H$-scaling method is better defined than when using the
$H$-scaling method. The lower panels show the resulting quality of
the collapsed wavelet coefficients as a function of time scale
for different distances $R$, using these best estimates of $p$ and $H$.}
\end{center}
\end{figure}

The determination of the exponents $p$ and $H$ is performed by
using the following steps when using the $H$-scaling law normalized curves:
\begin{enumerate}
\item We now re-organize all the $C_{a}(R)$ values by plotting
curves of $C_{a}(R)$ as a function of $R$ for various values of $a$.
These are now the $a$-curves.

\item The $a$-curves do not display any peculiar behavior with $R$.
They are monotonically increasing with $R$, and saturate at large
$R$ values. Among those $a$-curves, there is always a subset of curves
which are nearly parallel (or at least don't cross each other), which
is a behavior predicted by diffusive processes. If all the curves are
strictly parallel, one can conclude there is no underlying
diffusion at all. Other curves either cross this subset, or are simply
too noisy and are thus eliminated. This allows us to select the
$a$-curves we will use to invert for $p$ and $H$ using the $H$-scaling law.
Note that we will conserve ``raw'' $a$-curves, which can't be
approximated by any power-law.

\item Choose a trial $(p,H)$ and normalize each curve
according to these exponents and their respective $a$ and $R$ values
according to (\ref{mgmnsas}) ($H$-scaling law).
The normalized space scale axis is then re-sampled using
a geometrical sampling with a multiplicative factor of $1.01$, so that
all normalized $a$-curves are defined at common abscissae.

\item For each normalized value of the space scale, search for all normalized
$a$-curves which are
defined at that value. Let us assume there are $N_s$ such segments,
each one corresponding to a different $a$-curve (thus $a$ value).

\item Compute the average value of these normalized $a$-curves
that are defined at the same normalized time scale, as well as
the associated variance. If none or only one $a$-curve
is defined for the considered normalized space scale, the calculation is
not performed as the variance cannot be estimated. Then,
go to the next normalized time scale.

\item When all values of the normalized space scale have been considered,
compute the average of all the variances
that have been defined up to now. This average variance gives us an estimation
of how the various normalized curves collapse on top of each other in
the goal of defining a single master curve. It is approximately equivalent
to the square of the average width defined by superimposing all the 
normalized curves.

\item This algorithm is implemented using a $(p,H)$ grid, with $p$
varying from $0$ to $2$ by
steps of $0.01$, while $H$ varies within $[-1:1]$, with the same
step. When $H=0$, no
computation is made. A systematic search provides the couple of
exponents $(p,H)$ with the lowest average variance, i.e., the
best collapse of the wavelet coefficients as a function
of time scale and distance. In this case, there is no problem for $H=0$.
The top right panel of Figure
\ref{roundvalley} constructed for the Round Valley mainshock show the
contour lines representing equivalues of the average variance
(quantified in logarithmic scale in base $10$).

\item Once the best $(p,H)$ couple has been found, the
normalized wavelet coefficients $C_{a}(R)$ are calculated as a function of $R$
using the original non-normalized $a$-curves.
This is performed for the purpose of visualizing
the predicted collapse of the wavelet coefficients on the
real data-set, as shown in the right lower panel of Figure \ref{roundvalley}
constructed for the Round Valley mainshock.

\end{enumerate}

\end{article}


\begin{table}
\tablenum{1}
\caption{ Analysis of aftershock sequences of California
with the windowing method.
The first and second columns give the name and time of the mainshock.
$M$ is the mainshock magnitude, $T$ and $R$ are the temporal and spatial
windows used to select aftershocks, $M_0$ is the minimum magnitude of
aftershocks, $p$ is the Omori exponent measured over $t_{min}<t<T$,
$N$ is the number of aftershocks.
$H_r$, $H_a$ and $H_b$ are the diffusion exponents measured using
the average distance between aftershocks
and the barycenter, the large elliptical axis $a$,
and the short elliptical axis $b$ respectively.}
\begin{center}
\begin{tabular}{lccccccccccc}
\hline
earthquake & date  & $M$ & $T$ & $R$ & $M_0$ & $t_{min}$ & $N$ & $p$
  & $H_r$ & $H_a$ & $H_b$ \\
	& (dd/mm/yy) &	& (days) & (km) & & (days) & & & & & \\
\hline 
Kern-County      &21/07/52 &7.5 &5478&  70 &3.5 &1.0 &  281 &0.92 
&0.06& 0.05& 0.08\\
San Fernando     &09/02/71 &6.6 &1096&  40 &3.0 &0.1 &  274 &1.04 
&0.01& 0.03&-0.01\\
Oroville         &01/08/75 &5.7 &1826&  15 &2.0 &1.0 &  785 &1.09 
&0.04& 0.04& 0.04\\
Imperial Valley  &15/10/79 &6.4 &36  &  80 &2.5 &0.2 &  677 &1.44 
&0.41& 0.37& 0.19\\
Westmorland	 &26/04/81 &5.7 &  73&  20 &1.7 &0.2 &  587 &1.40 
&0.16& 0.16& 0.12\\
Coalinga         &02/05/83 &6.7 &1826&  22 &2.0 &1.0 & 3133 &1.03& 
0.04 &0.04& 0.01\\
Morgan-Hill      &24/04/84 &6.2 & 182&  30 &1.5 &0.02&  633 &0.60& 
0.11 &0.12& 0.44\\
Round-Valley     &23/11/84 &6.1 & 182&  15 &2.0 &0.1 & 1398 &0.94& 
0.09 &0.09& 0.11\\
North Palm Springs&8/07/86&5.6 & 365 &  15 &1.5 &1.0 & 2331 &1.10& 
0.04 &0.05& 0.04\\
Oceanside        &13/07/86 &5.4 &3650&  20 &2.0 &0.5 & 1926 &0.79& 
0.06 &0.08& 0.02\\
Chalfant Valley	 &21/07/86 &6.4 &1826&  20 &2.0 &1.0 & 2985 &1.16& 
0.06 &0.03& 0.16\\
Superstition-Hill&24/11/87 &6.6 &  18&  50 &1.8& 0.4 &  794 &1.21& 
0.20& 0.20& 0.10\\
Loma-Prieta	 &18/10/89 &7.0 &  36&  50 &2.0 &0.1 &  728 &1.05& 
0.11 &0.09& 0.29\\ 
Joshua-Tree 	 &23/04/92 &6.1 &  36&  30 &1.6& 3.0 & 3658 &1.11& 
0.09& 0.03& 0.27\\
Cape Mendocino	 &25/04/92 &6.5 &  36&  70 &2.0 &0.6 & 1197 &1.20& 
0.05 &0.01& 0.13\\ 
Landers     	 &28/06/92 &7.3 & 365&  60 &2.2& 3.0 & 7278 &1.05& 
0.00&-0.01& 0.02\\
Big Pine	 &17/05/93 &6.2 & 365&  25 &1.5 &2.0 &  780 &1.22& 
0.02 &0.00& 0.04\\ 
Northridge  	 &17/01/94 &6.7 &1826&  30 &2.0& 2.0 & 3254 &1.13& 
0.05& 0.06& 0.07\\
Nevada (Carter)	 &12/09/94 &5.5 & 365&  25 &2.5 &5.0 &  502 
&1.11&-0.01 &0.05& 0.00\\ 
Mammoth Lakes	 &15/05/99 &5.6 & 735&  10 &1.5 &0.2 & 1570 &0.84& 
0.09 &0.07& 0.16\\
Hector-Mine 	 &16/10/99 &7.1 &1826&  35 &2.5& 1.0 & 1812 &1.14& 
0.00&-0.01& 0.09\\
\\
\hline
\label{tabdifobs}
\end{tabular}
\end{center}
\end{table}

\begin{table}
\tablenum{2}
\caption{  Analysis of aftershock sequences of California
with the wavelet method.
The first column gives the name of the mainshocks. $p$ ($H$ method)
is the Omori's exponent obtained with the $H$-scaling law.
The quantities $a$ and $R$ are discussed in the text.
$p$ ($1/H$ method) is the Omori's exponent obtained with the
$1/H$-scaling law. $H$ ($H$ method) is the diffusion exponent
obtained with the $H$-scaling law. $H$ ($1/H$ method) is the diffusion
exponent obtained with the $1/H$-scaling law.}

\begin{center}
\begin{tabular}{lcccccc}
\hline
earthquake 		    &  $a$ range       & $R$ range        & 
$p$(``$1/H$'')	& $p$(``$H$'')      &  $H$(``$1/H$'')  & $H$(``$H$'') 
\\ 
& (days)& (km)& & & & 
\\ \hline 
Kern-County           &	$9.5-73$        & $40-150$       & $1.17$ 
		& $1.22$		  &  $-0.09$	   & $-0.07$ 
\\
San Fernando          &	$2-15$             & $5-30$            & $1.1$ 
& $^a$			  &  $0.02$		   & $^a$ 
\\
Oroville              &	$110-365$       & $4-30$         & $1.21$ 
		& $1.18$		  &  $0.03$		   & 
$0.03$           \\
Imperial Valley       &	$2-55$          & $41-200$       & $1.47$ 
		& $1.54$		  &  $-0.07$	   & $-0.07$ 
\\
Westmorland	 	    &	$0.75-75$       & $5-15$         & 
$1.43$  		& $0.22$		  &  $0.10$ 
	   & $0.52$           \\
Coalinga              &	$3-360$         & $3-20$         & $0.95$ 
		& $0.81$		  &  $0.08$		   & 
$0.11$           \\
Morgan-Hill           &	$0.75-256$      & $3-120$        & $0.51$ 
		& $0.57$		  &  $0.08$		   & 
$-0.02$          \\
Round-Valley          &	$1-30$          & $2-20$         & $0.72$ 
		& $0.35$		  &  $0.24$		   & 
$0.32$           \\
North Palm Springs    & $2.5-1825$     & $2-10$         & $1.14$ 
	      & $0.47$		  &  $0.01$		   & $0.26$ 
\\
Oceanside             &	$1-5.5$         & $5-20$         & $1.11$ 
		& $1.12$		  &  $0.02$		   & 
$0.01$           \\
Chalfant Valley	    &	$4-1460$       & $3-30$         & $1.15$ 
		& $1.12$		  &  $0.03$		   & 
$0.03$           \\
Superstition-Hill     &	$2-25$          & $5-40$         & $1.58$ 
		& $1.57$		  &  $0.06$		   & 
$0.04$           \\
Loma-Prieta	 	    &	$0.25-37$      & $7-200$        & 
$1.03$  		& $1.04$		  &  $-0.02$	   & 
$-0.01$          \\
Joshua-Tree 	    &	$7-26$          & $1.7-12$       & $0.98$ 
		& $0.93$		  &  $0.08$		   & 
$0.07$           \\
Cape Mendocino	    &	$3-128$         & $12-70$        & $1.11$ 
		& $1.12$		  &  $-0.01$	   & $-0.01$ 
\\
Landers     	    &	$15-1100$       & $3-180$        & $1.10$ 
		& $1.16$		  &  $-0.07$	   & $-0.06$ 
\\
Big Pine	 	    &	$3-365$         & $5-50$         & 
$1.11$  		& $1.10$		  &  $0.03$ 
	   & $0.02$           \\
Northridge  	    &	$15-730$        & $7-90$         & $1.28$ 
		& $1.31$		  &  $-0.03$	   & $-0.03$ 
\\
Nevada (Carter)       &	$18-90$         & $4-45$         & $1.19$ 
		& $1.37$		  &  $-0.15$	   & $-0.17$ 
\\
Mammoth Lakes	 	    &	$1-4$           & $2-8$         & 
$0.59$  		& $1.94$		  &  $0.20$ 
	   & $-0.72$          \\
Hector-Mine 	    &	$2-220$         & $2-30$         & $1.20$ 
		& $1.19$		  &  $-0.01$	   & $0.00$ 
\\
\\
\hline
\label{tab2wavelet}
\end{tabular}
\end{center}

\end{table}

{\small $^a$ The $H$ and $p$ values cannot be estimated with the $H$-wavelet
method for the San-Fernando sequence because there is no
clear minimum of the variance.}

\pagebreak

\begin{figure}
\psfig{file=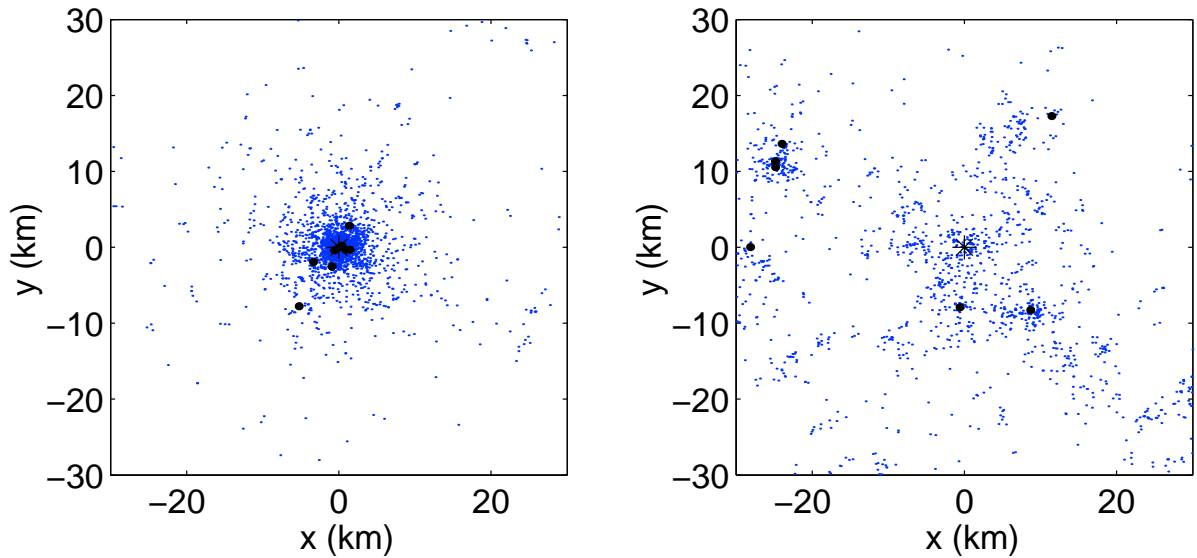,width=16cm}
\caption{ \label{map}
Maps of seismicity generated by the ETAS model with parameters
$b=1$, $\theta=0.2$, $\mu=1$, $d=1$ km, $\alpha=0.5$, $c=0.001$ day and a
branching ratio $n=1$. The mainshock occurs at the origin of space
with magnitude $M=7$ (black star). The minimum  magnitude is fixed at $m_0=0$.
The distances between mainshock and aftershocks follow a power-law
with parameter $\mu=1$ and the local Omori law is
$\propto 1/t^{1+\theta}$.  According to the theory developed
in the text, the average distance between the first mainshock and the
aftershocks is thus expected to grow as $R \sim t^{\theta/\mu}
\sim t^{0.2}$. The two plots are for different time periods of the
same numerical simulation, such that the same number of earthquakes
$N=3000$ is obtained for each graph: (a) time between $0$ and $0.3$ days;
(b) time between $30$ and $70$ yrs. At early times,
aftershocks are localized close to the mainshock, and then diffuse
and cluster around the largest aftershocks.}
\end{figure}
  \clearpage

\clearpage

\clearpage

\clearpage

\clearpage

\clearpage

\clearpage

\clearpage

\clearpage

\clearpage

\clearpage

\clearpage

\clearpage

\clearpage

\clearpage

\end{document}